\documentclass[fleqn,usenatbib]{mnras}

\usepackage{newtxtext,newtxmath}
\usepackage{multirow}
\usepackage{amsmath}

\usepackage[T1]{fontenc}

\DeclareRobustCommand{\VAN}[3]{#2}
\let\VANthebibliography\thebibliography
\def\thebibliography{\DeclareRobustCommand{\VAN}[3]{##3}\VANthebibliography}

\usepackage{graphicx}	
\usepackage{amsmath}	
\usepackage[utf8]{inputenc}
\usepackage[export]{adjustbox}
\usepackage{wrapfig}
\usepackage{ulem}

\title[Multi-shock scenario]{Radio relics radio emission from multi-shock scenario}

\author[G. Inchingolo et al.]{
Giannandrea Inchingolo,$^{1,2}$\thanks{E-mail: giannandr.inchingolo@unibo.it (GI)}
D. Wittor,$^{3,1}$
K. Rajpurohit$^{1,2,4}$
F. Vazza$^{1,2,3}$
\\
$^{1}$Dipartimento di Fisica e Astronomia, Università di Bologna, Via Gobetti 92/3, 40129, Bologna, Italy\\
$^{2}$Institute of Radioastronomy - INAF, Via Gobetti 101, 40129 Bologna, Italy\\
$^{3}$Hamburger Sternwarte, Gojenbergsweg 112, 21029 Hamburg, Germany\\
$^{4}$Th\"{u}ringer Landessternwarte, Sternwarte 5, 07778 Tautenburg, Germany
}

\date{Accepted October 21st, 2021. Received July 5th, 2021}

\pubyear{2021}

\begin{document}
\label{firstpage}
\pagerange{\pageref{firstpage}--\pageref{lastpage}}
\maketitle

\begin{abstract}
Radio relics are giant ($\sim$Mpc) synchrotron sources that are believed to be produced by the (re)acceleration of cosmic-ray electrons (CRe) by shocks in the intracluster medium.
In this numerical study, we focus on the possibility that some radio relics may arise when electrons  undergo diffusive shock acceleration at multiple shocks in the outskirts of merging galaxy clusters. This multi-shock (MS) scenario appears viable to produce CRe that emit visible synchrotron emission.
We show that electrons that have been shocked multiple times develop an energy spectrum that significantly differs from the power-law spectrum expected in the case of a single shock scenario.
As a consequence, the radio emission generated by CRe that shocked multiple times is higher than the emission produced by CRe that are shocked only once. In the case explored in this paper, the radio emission produced in the two scenarios differ by one order of magnitude.
In particular in the MS scenario, the simulated relic follows a KGJP spectral shape, consistent with observation.
Furthermore, the produced radio emission is large enough to be detectable with current radio telescopes (e.g. LOFAR, JVLA).
\end{abstract}

\begin{keywords}
galaxies:clusters:general
\end{keywords}



\section{Introduction}
\label{sec:intro}
Radio relics usually located in the outskirts of merging galaxy clusters are giant ($\sim$Mpc) synchrotron sources that are believed to be produced by cosmic-ray electrons (CRe) (re)-accelerated by merger induced shock waves in the intracluster medium (ICM; \citet{ensslin1998, 1999ApJ...518..594R, bj14, vanweeren2019review, 2019SSRv..215...14B}).
The connection between shocks and relics has been confirmed by finding the surface brightness and temperature discontinuities in the X-ray observations at the location of relics \citep[e.g.][]{2008A&A...486..347G, 2013PASJ...65...16A,vanweeren2019review}.

The details of the acceleration mechanisms in radio relics are still not fully understood. The widely accepted mechanism for acceleration of relativistic cosmic-ray (CR) particles at shock fronts is diffusive shock acceleration (DSA) \citep[e.g.][]{1987PhR...154....1B}.
DSA is based on the original idea of \citet{1949PhRv...75.1169F}, according to which particles are scattered upstream and downstream of the shock by plasma irregularities, gaining energy at each shock crossing.
In recent years, deep X-ray observations performed with \textit{Chandra}, \textit{XMM-Newton}, and \textit{Suzaku} have led to an increase in the number of shocks detected in merging galaxy clusters \citep[e.g.][ for recent works]{2017A&A...600A.100A, 2017MNRAS.464.2896C, 2018MNRAS.476.5591B}.
Radio and X-ray observations suggest that radio relics probe particle acceleration by weak shocks, $\mathcal{M} \leq 5$, \citep[e.g.][]{2009A&A...494..429B,vw10,2012MNRAS.426...40B,Hoang2017sausage,2017A&A...600A.100A,2018ApJ...852...65R,2018MNRAS.476.5591B, 2019ApJ...873...64D} in a high-$\beta$ ($\beta = P_{th}/P_B$, i.e., the ratio between the thermal and magnetic pressures) environment such as the ICM, where the thermal pressure dominates over the magnetic pressure.
However, X-ray and radio estimates of shocks' strength are typically in disagreement, possibly because these two proxies probe different parts of the underlying Mach number distribution \citep[e.g. see][for a recent discussion of this issue]{2021arXiv210608351W}.
In the weak shock regime, the acceleration efficiencies of cosmic-ray protons (CRp) are poorly understood, although current models and simulations predict acceleration efficiencies (defined as the ratio between the shock kinetic power and the energy flux of accelerated cosmic rays) that are less than a few percent \citep[e.g.][]{2018ApJ...864..105H, 2019ApJ...883...60R, 2020ApJ...892...86H, 2020MNRAS.495L.112W}, in agreement with direct constraints coming from $\gamma$-ray non-detections of galaxy clusters \citep[e.g.][for review]{ack10, ackermann14, ackermann16, 2021NewA...8501550W}.
On the other hand, the observed connection between radio relics and shocks in merging galaxy clusters demonstrates that the electron acceleration (or re-acceleration) at these shocks is efficient, in the sense that even weak shocks ($\mathcal{M} \leq 2$) are associated with detectable radio emission. This implies a surprisingly large ratio of electron-to-proton CR acceleration efficiencies for DSA, because at the same time CR protons have never been detected in the ICM  \citep[e.g.][]{va14relics, bj14,2015MNRAS.451.2198V, scienzo16,2020A&A...634A..64B}.

Even if radio power of some relics can be explained by the acceleration of electrons from the thermal pool (i.e. DSA mechanism) \citep{locatelli2020dsa}, this mechanism alone cannot explain the high radio power of the majority of relics \citep{2016MNRAS.460L..84B, 2016MNRAS.461.1302E, Hoang2017sausage}. To mitigate the problem of the high acceleration efficiencies implied by weak cluster shocks, recent theoretical models assume a pre-existing population of CRe at the position of the relic that is re-accelerated by the passage of the shock \citep[e.g.][]{2005ApJ...627..733M, 2011ApJ...728...82M, kr11, ka12, 2014ApJ...788..142K, 2013MNRAS.435.1061P, 2020A&A...634A..64B}. This would soften the above theoretical problems, because  the population of CRs processed by $\mathcal{M} \leq 3-4$ shocks is predicted to be dominated by the re-accelerated fossil populations, and not the freshly accelerated one.    
The re-acceleration scenario is supported by the observation of radio galaxies located nearby or within a few radio relics \citep[e.g.][]{2014ApJ...785....1B, 2015MNRAS.449.1486S, 2016MNRAS.460L..84B, 2017NatAs...1E...5V, digennaro2018saus}.
However, it is not obvious that the injection of fossil electrons by one or a few radio galaxies, can automatically produce a uniform population of electrons, capable of producing the high degree of coherence of the radio emission observed in a few giant relics: in radio relics like "the Sausage" and "the Toothbrush" the spectral properties of the emission are very coherent across $\sim 2$ $\rm Mpc$,  requiring a very uniform distribution of coeval fossil electrons in the shock upstream \citep[e.g.][]{2010Sci...330..347V,2016ApJ...818..204V,2018ApJ...852...65R,rajpurohit2020toothbrush, digennaro2018saus}. 

Complementary to the above scenario, we thus focus here on a specific mechanism potentially alleviating this problem, i.e. we consider a multiple-shock (MS) scenario in which multiple, wide merger shocks sweeping the ICM in sequence can produce a large-scale and uniform distribution of mildly relativistic electrons. 
A similar mechanism has been very recently analyzed  by \citet{kang2021diffusive}, in the context of the acceleration of cosmic ray protons via DSA. 
The existence of multiple populations of shock waves sweeping the ICM along a variety of angles with respect to the leading axis of mergers, and  possibly merging into larger shocks have been recently explored by several simulations  \citep[e.g.][]{2015ApJ...812...49H,2020MNRAS.498L.130Z,2021MNRAS.501.1038Z}

In our work, we focus on MS electron acceleration and the radio emission generated in this way.
In detail, we analyse the simulation of a massive $M_{200} \approx 9.7 \cdot 10^{14} \ \mathrm{M}_{\odot}$ galaxy cluster from $z=1$ to $z=0$ \citep{2017MNRAS.464.4448W}, and we compute the radio emission generated by particles following the merging of the cluster, showing that MS electrons develop, on average, large enough radio emission to be detectable with current radio telescopes (e.g. LOFAR).

This paper is organized as follow.
In Section \ref{sec:method} we describe the numerical set-up used for the galaxy cluster simulation and the model used to simulate the evolution of electron spectra.
The results obtained are analyzed in Section \ref{sec:families}, where we distinguish two different relics probed by the particle simulated and we study the radio emission of these particles catalogued by their shock history.
The detailed study of the integrated radio emission is presented in Section \ref{sec:integrated_radio}.
Section \ref{sec:conclusion} summarizes the results obtained in the paper and discusses future work.

\section{Numerical Method}
\label{sec:method}

\subsection{Simulation setup}\label{subsec:crater}

In this work, we study a massive, $M_{200} \approx 9.7 \cdot 10^{14} \ \mathrm{M}_{\odot}$, galaxy cluster that was simulated and analysed in \citet[][]{2016Galax...4...71W,2017MNRAS.464.4448W}.
This cluster is interesting for a comparison with real observations as it undergoes a major merger at redshift $z \approx 0.27$, producing detectable giant radio relics.
The cluster was simulated with the cosmological magneto-hydrodynamic (MHD) code \textsc{ENZO} \citep{ENZO_2014} and analysed with the Lagrangian tracer code Cosmic-Ray Tracers (\textsc{CRaTer}) \citep{2017MNRAS.464.4448W}.
In the following, we give a brief overview on the used simulation setup.
For specific details, we point to section 2.1 in \citet{2017MNRAS.464.4448W}.

The \textsc{ENZO} code follows the dark matter using a N-body particle-mesh solver \citep{1988csup.book.....H} and the baryonic matter using an adaptive mesh refinement (AMR) method \citep{1989JCoPh..82...64B}.
More specifically, \citet{2017MNRAS.464.4448W} used the piecewise linear method \citep{1985JCoPh..59..264C} in combination with the hyperbolic Dedner cleaning \citep{2002JCoPh.175..645D}. The simulation covers a root grid with a comoving volume of $\sim (250 \ \mathrm{Mpc})^3$ sampled with $256^3$ grid cells and dark matter particles.
An additional comoving volume of size $\sim (25 \ \mathrm{Mpc})^3$ has been further refined using 5 levels of AMR, i.e. $2^5$ refinements, for a final resolution of $31.7 \ \mathrm{kpc}$.
The chosen AMR criteria, i.e. based on the over-density and the 1D velocity jump, ensure that about $\sim 80 \ \%$ of the cluster volume are refined at the highest AMR level.

We study this cluster in detail because it is a massive one, it has been already the subject of several works by our group, and because the fairly large dynamical range and number of available snapshots is optimal for our analysis involving tracer particles (see below). However, the final magnetic field reached through small-scale dynamo amplification in this object is kept artificially small by the spatial resolution, which is not enough to ensure a large enough Reynolds number to enter an efficient small-scale dynamo amplification regime, as studied in \citet{va18mhd}. Therefore, for simulating the injection and advection of CRe in this system, we re-normalized the magnetic field strength, measured by the tracers,by a factor 10. The re-normalization results typical magnetic field strengths of $\sim 0.1-0.2$ $\rm \mu G$ in our relics.
In fact, the electron cooling depends rather weakly on the renormalization of magnetic field strengths, because inverse Compton cooling dominates over synchrotron cooling (see the denominator in Eq. \ref{eq:ic}.

Using \textsc{CRaTer}, \citet{2017MNRAS.464.4448W} used a total of $\sim 1.3 \cdot 10^7$ Lagrangian tracer particles to analyse the cluster's evolution between $z = 1$ and $z = 0$, at a (nearly constant) time resolution of $\Delta t=31 \rm ~Myr$.
Following the gas distribution of the ICM, \textsc{CRaTer} injects particles with a fixed mass, i.e. in our case $m_{\mathrm{tracer}} \approx 10^8 \ \mathrm{M}_{\odot}$, into the simulation.
The tracers' velocities are computed by interpolating the local velocities to the tracers' position using a \textit{cloud-in-cell}-interpolation method.
An additional velocity correction term was used in \citet{2017MNRAS.464.4448W} to account for mixing motions that might be underestimated in the case of complex flows \citep{Genel_2014_following_the_flow}.
The velocity interpolation schemes have been extensively tested in \citet{2017MNRAS.464.4448W} and \citet{wittorPHD}.

The tracer particles use a temperature-based shock finder to detect shocks in the ICM.
The corresponding Mach number is computed from the Rankine-Hugoniot relation, assuming $\gamma = 5/3$, as

  \begin{align}
  M = \sqrt{\frac{4}{5} \frac{T_{\mathrm{new}}}{T_{\mathrm{old}}} \frac{\rho_{\mathrm{new}}}{\rho_{\mathrm{old}}} + \frac{1}{5}}.
 \end{align}
 
 Here, $T$ and $\rho$ are the temperature and density in the pre- and post-shock.
 
 We have specifically chosen the cluster simulation presented in \citet[][]{2016Galax...4...71W,2017MNRAS.464.4448W} for our analysis. \citet[][]{2017MNRAS.464.4448W} found that a significant fraction of the particles that produce giant radio relics at $z \approx 0$, have crossed several shocks before, see figure 12 and Section 3.5 therein. 
 Hence, the radio emitting particles should have been subjected to several cycles of shock (re-)acceleration, making this simulation a perfect candidate for our analysis.

\begin{figure*}
    \centering
    \includegraphics[width=1\textwidth]{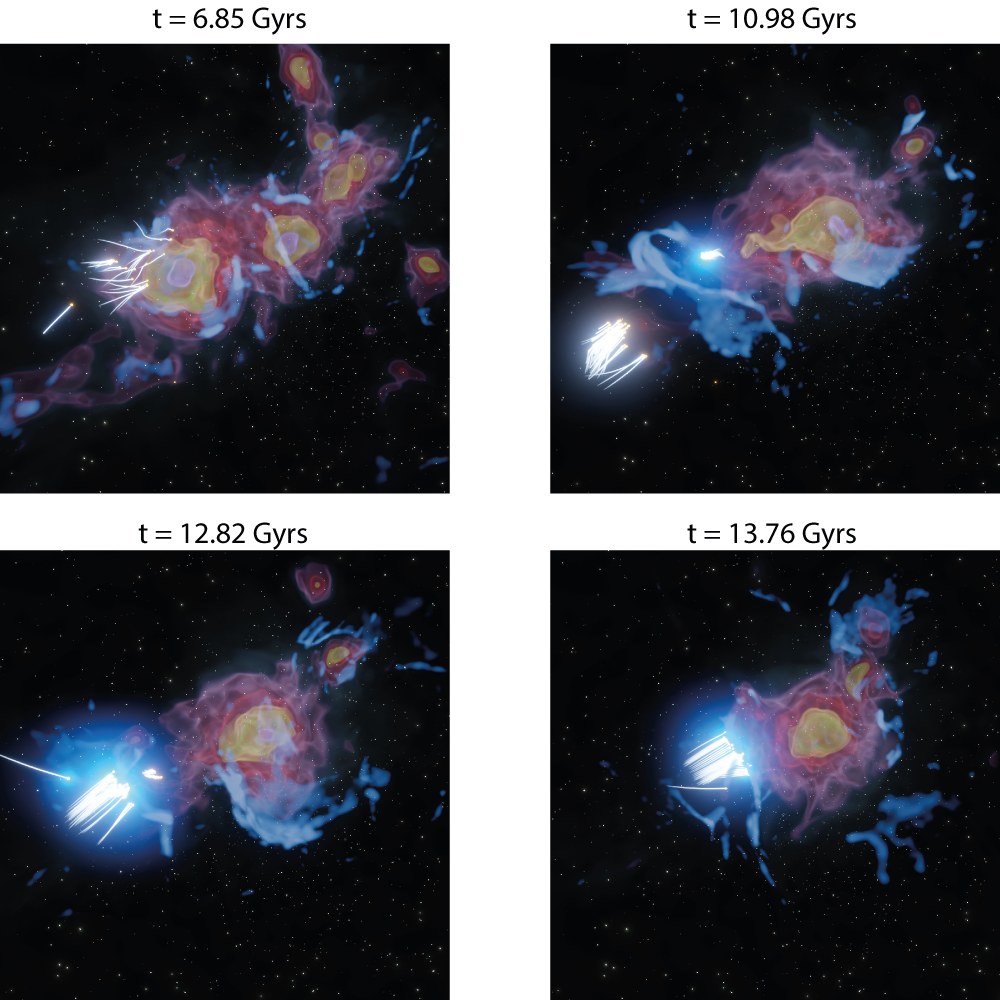}
    \caption{Snapshot sequence at different times of, respectively, the galaxy cluster baryonic density (purple-yellow), the radio emission at $1.4$ GHz (light blue) and the tracers with their path. The tracers change from yellow (not active) to blue (active) when they cross a shock front in the simulation.}
    \label{fig:video}
\end{figure*}
We use a 3D rendering of this merger event to better describe the sequence of mergers (leading to multiple shock waves) which interest a particular sector of the cluster.
Figure \ref{fig:video} shows a snapshot sequence obtained from a cinematic scientific visualization realized from the simulation data\footnote{The video is called "\textit{The VLA shedding lights on the origin of radio relics}" and it was recently awarded the $1^{st}$ prize for the NRAO Image Contest for the celebration of VLA $40^{th}$ anniversary. The video is available at the following link: \hyperlink{https://vimeo.com/464248944/3fc17a5b8b}{https://vimeo.com/464248944/3fc17a5b8b}.}.
The video shows the baryonic density (purple-yellow) surrounded by the volumetric radio emission at $1.4$ GHz (light blue) during the formation of the galaxy cluster.
The tailed spheres highlight the evolution of a selection of tracers. 
Initially, all the tracers are yellow and when they cross a shock front, they are activated, changing color to bright blue.
The sequence in Fig. \ref{fig:video} shows the evolution of two streams of tracers (``beam").
This qualitative analysis of the cluster merging evolution shows an history of MS scenario before tracers arrive at the end of the simulation.
In this paper, we analyse the spectral evolution measured by the tracers in this simulation.

\subsection{Simulating the evolution of electron spectra}
\label{subsec:fokker}

We solve the time-dependent diffusion-loss equation of relativistic electrons represented by tracer particles, using the standard \citet{1970JCoPh...6....1C} finite-difference scheme implemented in a serial code written in  IDL language.
We used $N_{\rm b}=10^5$ equal energy bins in the $\gamma_{\rm min} \leq \gamma \leq \gamma_{\rm max}$ Lorentz factor, with  $\gamma_{\rm min}=1$ and $\gamma_{\rm max}=4.5\times10^5$ (hence $\rm d\gamma=5$). The code we used to evolve our particle spectra is freely available \footnote{\url{https://github.com/FrancoVazza/IDL_FP}}.

We are concerned with the evolution of relativistic electrons injected and/or re-accelerated by shocks, at the periphery of clusters and on timescales of a few Gigayears ($\leq 3 \rm ~Gyr$). 
For this specific task, we only have to evolve the energy spectra for $ \sim 7000$ tracers, necessary to sample the spatial extension of radio relics formed in the system by $z \approx 0$. 
The combination of the limited amount of tracers and of the relatively small number of snapshots to process (up to 238) allowed us to resort to the serial implementation of the Fokker Planck solver already used in previous work \citep[e.g.][]{rajpurohit2020toothbrush}.
Notice that, unlike the more recent work presented in \citet{2021arXiv210204193V}, in this implementation, we evolve the electron spectra in $\gamma$ space, and not in momentum space.
This introduces a small error in the low energy part of the spectra, where the injected distribution from shock acceleration is a power-law in momentum space, but not in $\gamma$ space (since of course $E^2=m^2c^4+ p^2 c^2$).
The ultra-relativistic simplification used here is however suitable when focusing on the radio emitting electrons ($\gamma \geq 10^2-10^3$) and also considering that the accumulated particle population at low energy is small in the short time range considered \citep[e.g.][]{sa99}.

We considered a reduced Fokker-Planck equation without injection and escape terms (i.e. Liouville equation), and neglected the spatial diffusion of cosmic rays (which is appropriate for the $\sim \rm MeV-GeV$ electrons considered in this work),  which allows us to track the evolution of the number density of relativistic electrons as a function of their energy, $N(\gamma)$, computed separately for each tracer particle:

\begin{equation}
    {\frac{\partial N}{\partial t}}
    =
    {\frac{\partial}{\partial \gamma}} \left[
    N \left( \left|{\frac{\gamma}{\tau_{\rm rad}}}\right| + \left|{\frac{\gamma}{\tau_{\rm c}}}\right| +
    {\frac{\gamma}{\tau_{\rm adv}}} - \left|{\frac{\gamma}{\tau_{\rm acc}}}\right| \right) \right],
    \label{eq11}
\end{equation}

We use the approximation 

\begin{equation}
    \dot{\gamma} \approx  \left|{\frac{\gamma}{\tau_{\rm rad}}}\right| + \left|{\frac{\gamma}{\tau_{\rm c}}}\right| +
    {\frac{\gamma}{\tau_{\rm adv}}} - \left|{\frac{\gamma}{\tau_{\rm DSA}}}\right| ,
    \label{eq12}
\end{equation}

where $\tau_{\rm rad}$, $\tau_{\rm c}$, and $\tau_{\rm adv}$ are respectively the loss timescales for the radiative, Coulomb and expansion (compression) processes that we define in Sec. \ref{sec:lossterm}.
$\tau_{\rm DSA}$ represents instead the acceleration timescale due to DSA that we estimate in Sec. \ref{sec:DSAacc}.

The numerical solution is obtained using the \citet{1970JCoPh...6....1C} finite difference scheme: 

\begin{equation}
N(\gamma,t+dt)=\frac{{{N(\gamma,t)}/{dt}} + N(\gamma+d\gamma,t+dt){{\gamma}}} 
{1/dt + {{\gamma}}/d\gamma} + Q_{\rm inj}(\gamma)  ,
    \label{eq13}
\end{equation}
where in the adopted  splitting-scheme to perform the finite differences we assumed $N(\gamma +d\gamma/2)=N(\gamma+d\gamma)$ and $N(\gamma-d\gamma/2)=N(\gamma)$, where $Q_{\rm inj}$ accounts for the injection by shocks. The latter isis regarded as an almost instantaneous process, considering that timescales are much shorter than the time step of our integration, $\delta t \approx 31 \rm Myr$ (see Eq. \ref{eq:tDSA} below).

\subsubsection{Loss Terms}
\label{sec:lossterm}

The timescales associated to the energy losses by  radiative, Coulomb and expansion (compression) processes are  given by the following formulae, adapted from \citet{bj14}: 

\begin{equation}
    \tau_{\rm rad} =\frac
    {7720 \rm ~Myr} {(\gamma/{300})\left[\left(\frac{B}{3.25 \rm \mu G}\right)^2 + (1+z)^4\right]} , 
    \label{eq:ic}
  \end{equation}

\begin{equation}
    \tau_{\rm c} =
    7934 \rm ~Myr \left\{ {{\frac{n/10^{-3}}{{\gamma/300}}}}
    \left( 1.168 + {\frac{1}{75}}ln \left( {\frac{\gamma/300}{ n/10^{-3} }} \right) \right) 
\right\}^{-1}    
\label{eq:coulomb}
\end{equation}

and

\begin{equation}
    \tau_{\rm adv} = \frac{951 \rm ~Myr}{ 
    \nabla \cdot \mathbf{v}/10^{-16}} ,
\label{eq:adv}
\end{equation}

\noindent
in which where the density $n$ is measured in [$\rm cm^{-3}$], $B$ in [$\rm \mu G$] and the gas divergence is measured in $\nabla \cdot \mathbf{v}$ in [$1/\rm s$]. Bremsstrahlung losses can be safely neglected in this case, because for the typical ICM conditions encountered also their timescale is much larger than the ones of all other loss channels. Inverse Compton and synchrotron losses are by far the most relevant for the evolution of electrons considered in this work, owing to their peripheral location and low gas density.

\subsubsection{Shock (re-)acceleration}
\label{sec:DSAacc}

Predicting the spectrum of injected "fresh" relativistic electrons injected by weak shocks, as well as their spectrum after shock reacceleration, is far from being a solved problem. In this paper we follow a relatively simple approach, motivated by the existing literature on the subject and meant to simplify the steps to determine the post-shock spectrum of radio emitting electrons. 

We rely here on the DSA model by \citet{kr11}, which assumes that the injection Lorentz factor of electrons is related to the injection momentum ($\gamma_{\rm inj}=\sqrt{1+p^2_{\rm inj}/m_e^2c^2}$), where $p_{\rm inj}$ in DSA is assumed to be a multiple of the thermal momentum of {\it protons}, i.e. $p_{\rm inj}= \xi p_{\rm th}$ ($p_{\rm th}=\sqrt{2 k_b T_d m_p}$, where $k_b$ is the Boltzmann constant). Following \citet{kr11}, we compute $\xi$ based on the fit formula given from their one-dimensional convection-diffusion simulations:  

\begin{equation}
\xi_{\rm inj}=1.17 \frac{m_p v_d}{p_{\rm th}} \cdot (1+\frac{1.07}{\epsilon_B})\frac{\mathcal{M}^{0.1}}{3^{0.1}})    
\end{equation}
where $v_d$ is the downstream shock velocity and $\epsilon_B$ is the ratio of magnetic field strength between the $B_0$ downstream magnetic field generated by the shock, and $B_\perp$ is the magnetic field perpendicular to the shock normal.
 We set here $\epsilon_B=0.23$ \citep[][]{2013MNRAS.435.1061P} and obtaining values in the range $\xi_{\rm inj} \sim 2.5-3.5$ and $\gamma_{\rm inj} \sim 10-20$ for our shocks. 

The source term for relativistic electrons in Eq.\ref{eq13} assumes an energy distribution that follows a power-law \citep[e.g.][]{1962SvA.....6..317K,sa99}:

\begin{equation}
    Q_{\rm inj}(\gamma) = K_{\rm inj,e} ~\gamma^{-\delta_{\rm inj}} \left(1-\frac{\gamma}{\gamma_{\rm cut}}\right)^{\delta_{\rm inj}-2} ,
    \label{eq:xi}
\end{equation}
in which the initial slope of the input momentum spectrum, $\delta_{\rm inj}$, is computed based on the standard DSA prediction, i.e. $\delta_{\rm inj} = 2 (\mathcal{M}^2+1)/(\mathcal{M}^2-1)$.  
The cuff-off energy, $\gamma_{\rm cut}$,  is the defined for every shocked tracer as the maximum energy, beyond which the radiative cooling timescale is shorter than the acceleration timescale, $\tau_{\rm DSA}$:

\begin{equation}
\tau_{\rm DSA} = \frac{3~D(E)}{V_s^2} \cdot \frac{r(r+1)}{r-1} ,
\label{eq:tDSA}
\end{equation}
in which  $r$ is the shock compression factor, $V_s$ is the shock velocity, and $D(E)$ is the  diffusion coefficient of relativistic electrons, as a function of their energy \citep[e.g.][]{gb03}.
The specific energy-dependent value of $D(E)$ is little constrained because it depends on the local conditions of the turbulent plasma, and it is critical to limit the maximum energy in DSA  \citep[e.g.][]{ka12}. However, the latter 
is not an issue for our simulation, because all plausible choices of $D(E)$ in Eq.~\ref{eq:tDSA} give an acceleration timescale many orders of magnitude smaller than the typical cooling time of radio emitting electrons, whose energy distribution  be assumed to follow a power law within the energy range of interest,  at least the moment of .their injection.  We can set therefore $\gamma_{\rm cut} = \gamma_{\rm max}$ in this work.

This also motivates the fact that we can model shock injection by DSA by adding the newly created population of particles across timesteps (see Eq.~\ref{eq13} above), without integrating a source term as needed for the much slower re-acceleration by turbulence (see below).  

 Under these assumptions, the rate of injection of relativistic electrons in the downstream is:
 \begin{eqnarray}
K_{\rm inj,e}= 4 \pi ~ K_{e/p} \int_{p_{\rm inj}}^{p_{\rm cut}} (\sqrt{p^2+1}-1) f_N ~p^{-(\delta_{\rm inj}+2)} \cdot \nonumber
\\
\cdot \exp[-(p/p_{\rm cut})^{2}]  ~p^2 dp  ~dx_t^2  ~V_s  ~dt 
\label{eq:phicr}
\end{eqnarray}
with
\begin{equation}
f_N = \frac{n_d}{\pi^{3/2}}p_{\rm th}^{-3} \exp{(-\xi_{inj}^2)}
 \end{equation}
 
and where $K_{e/p}$  is the  electron-to-proton ratio. 
Following \citet{2020JKAS...53...59K} we use $K_{e/p}=(m_p/m_e)^{(1-\delta_{\rm inj})/2}$,  which gives $K_{e/p} \sim 10^{-2}$ for an injection spectral index of $\delta_{\rm inj} \approx 2.3$, in line with the injection spectral index of local Galactic supernova remnants \citep[e.g.][]{2007Natur.449..576U}.
 
$dx_t^2$ is the surface element associated to each shocked tracer particle, and is computed considering that $dx_t^3 = dx^3/n_{\rm tracers}$ is the initial volume associated to every tracer at the epoch of their injection ($n_{\rm tracer}$ being the number of tracers in every cell) and $dx_t(z)^3=dx_t^3 \cdot \rho_t/\rho(z)$ is the relative change of the volume associated to each tracer as a function of $z$, based on the ratio between the density at injection, $\rho$ and the density of cells where each tracer sits as a function of redshift, $\rho(z)$.

This procedure allow us to guess the acceleration efficiency of relativistic electrons at the shock, at least to a first degree of approximation and with a modest computing time. Of course, the physical uncertainty behind this is of course very large, and dedicated simulations are needed to fully solve the acceleration cycle of relativistic electrons by weak merger shocks,  for the possible range of shock obliquities and typical plasma conditions of the ICM \citep[][]{Guo_eta_al_2014_II,2015PhRvL.114h5003P,2019ApJ...876...79K,2020ApJ...897L..41X,2021ApJ...915...18H}.
\bigskip

Beside the {\it direct} injection of relativistic electrons by shocks, we also include the effect of shock {\it re}-acceleration on existing relativistic electrons \citep[e.g.][]{2005ApJ...627..733M,kr11,ka12}.
According to DSA, the input particle spectrum, $N_0(x)$, becomes

\begin{eqnarray}
N(\gamma)=(\delta_{\rm inj}+2) \cdot \gamma^{-\delta_{\rm inj}} \int_{\gamma_{min,re}}^\gamma N_0(x) x^{\delta_{\rm inj}+1} dx ,
\label{eq:shock_re}
\end{eqnarray}
where $\delta_{\rm inj}$ is the local slope within each energy bin.
We consider that the minimum momentum for the electron re-acceleration by shocks is the injection momentum $p_{\rm inj}$, above which DSA is expect to operate \citep{2020JKAS...53...59K}. So we set $\gamma_{min,re-e} = \gamma_{\rm inj}$as a lower bound for the integration in Eq. \ref{eq:shock_re}.

\section{MS scenario of electrons re-acceleration}
\label{sec:families}
In this section, we analyze the properties of tracer particles used to probe the evolution of the simulation, focusing on their shock time.
We select more than 7000 tracers that cross shocks with Mach number $M\geq2$ during the entire evolution of the simulation.
By construction, all these tracers cross a shock at the end of the simulation $t_{\mathrm{end}}=13.76$ Gyrs.
We investigate if these tracers crossed other shocks before the final one and if, how many times.
We divide the tracers in different families according to the number of shocks they cross during the simulation evolution. 
Tracers of Family 1 are only accelerated by the shock at the end of the simulation.
Family 2, 3 and 4 have been shocked respectively one, two and three times before they cross the final shock.
Details of the families population are collected in Tab. \ref{tab:families_stat}.
\begin{table}
    \centering
    \begin{tabular}{c|cc|cc}
         & Relic A & & Relic B & \\
    \hline     
    Family 1 & 1297 & 91.92\% & 1709 & 28.91\% \\
    Family 2 & 114 & 8.08\% & 3687 & 62.36\% \\
    Family 3 & & & 489 & 8.27\% \\
    Family 4 & & & 27 & 0.46\% \\
    Total & 1411 & & 5912 & 
    \end{tabular}
    \caption{Tracer population statistic in the different families for both Relic A and Relic B.}
    \label{tab:families_stat}
\end{table}

\begin{figure}
    \centering
    \includegraphics[width=\columnwidth]{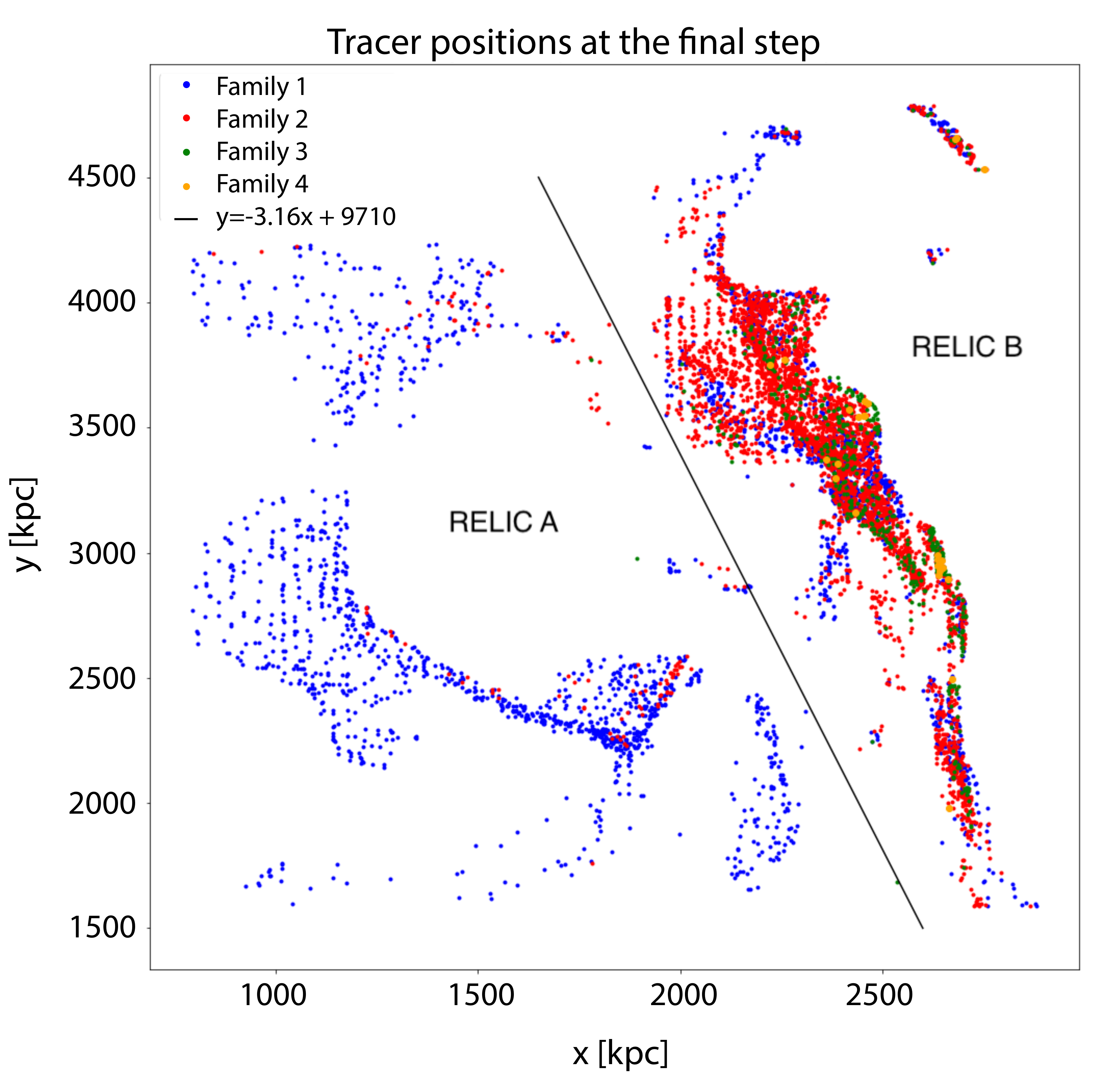}
    \caption{2D map of the tracer position at the final step. We distinguish two relics for the analysis. The black oblique line (defined by the equation in the legend) divides the region of the two relics, namely Relic A (left) and Relic B (right)}
    \label{fig:tracer_pos}
\end{figure}
Figure \ref{fig:tracer_pos} shows the $(x,y)$ projection of Family 1 (blue), Family 2 (red), Family 3 (green), and Family 4 (orange) tracers at $t_{\mathrm{end}}$. 
According to these positions, we divide the tracers in two groups named "Relic A" and "Relic B" in Fig. \ref{fig:tracer_pos}.
We observe that Relic A is composed mostly by Family 1 tracers, with the presence of $8 \ \%$ of Family 2 tracers.
Relic B, instead, is composed of  more than $62 \ \%$ by Family 2 tracers and for $\sim28 \ \%$ by Family 1 tracers, with the presence of families with higher number of shocks, as reported in Tab. \ref{tab:families_stat}.
As first approach, also motivated by the fact that the differences in the timing of shocks within each family of electrons is typically $\leq \rm ~Gyr$, we computed the energy evolution of each family based on the family-averaged fields, i.e. assuming at each timestep that the entire family of particles is charactersied by the same values of density, temperature and magnetic field, and that all particles in the same family are shocked at the same time. 
For this family-averaged analysis, we chose the shock times of each family as the ones at which the majority of the tracers cross a shock simultaneously.
This is of course a gross approximation, but it is enough to allow us to obtain some first important information on the electron energy spectrum based on the MS scenario and the subsequent radio emission.
A detailed report of the family-averaged approach is available as Appendix of this paper \ref{sec:appendix}.

\subsection{Relic A}
For Relic A, the family-averaged quantities are collected in Tab. \ref{tab:relicA_stat}. 
We use these quantities to compute the time evolution of the electron energy  spectrum according to the model introduced in Sec. \ref{subsec:fokker}.
\begin{table}
    \centering
    \begin{tabular}{c|c|c|c|c|c}
         & Time [Gyrs] & Mach & B [$\mu$G] & $\rho$ [g/cm$^{-3}$] & T [K]\\
    \hline
    Family 1 & & & & \\
    Shock 1 & 13.76 & 2.6 & $1.7\times10^{-1}$ & $1.3\times10^{-28}$ & $2.5\times10^{7}$ \\
    \hline
    Family 2 & & & & \\
    Shock 1 & 12.69 & 3.8 & $1.1\times10^{-1}$ & $1.6\times10^{-28}$ & $2.7\times10^{7}$ \\
    Shock 2 & 13.76 & 2.3 & $2.2\times10^{-1}$ & $2.0\times10^{-28}$ & $3.1\times10^{7}$
    \end{tabular}
    \caption{Family-averaged quantities at selected shock times for families in Relic A.}
    \label{tab:relicA_stat}
\end{table}

\begin{figure}
    \centering
    \includegraphics[width=0.75\columnwidth]{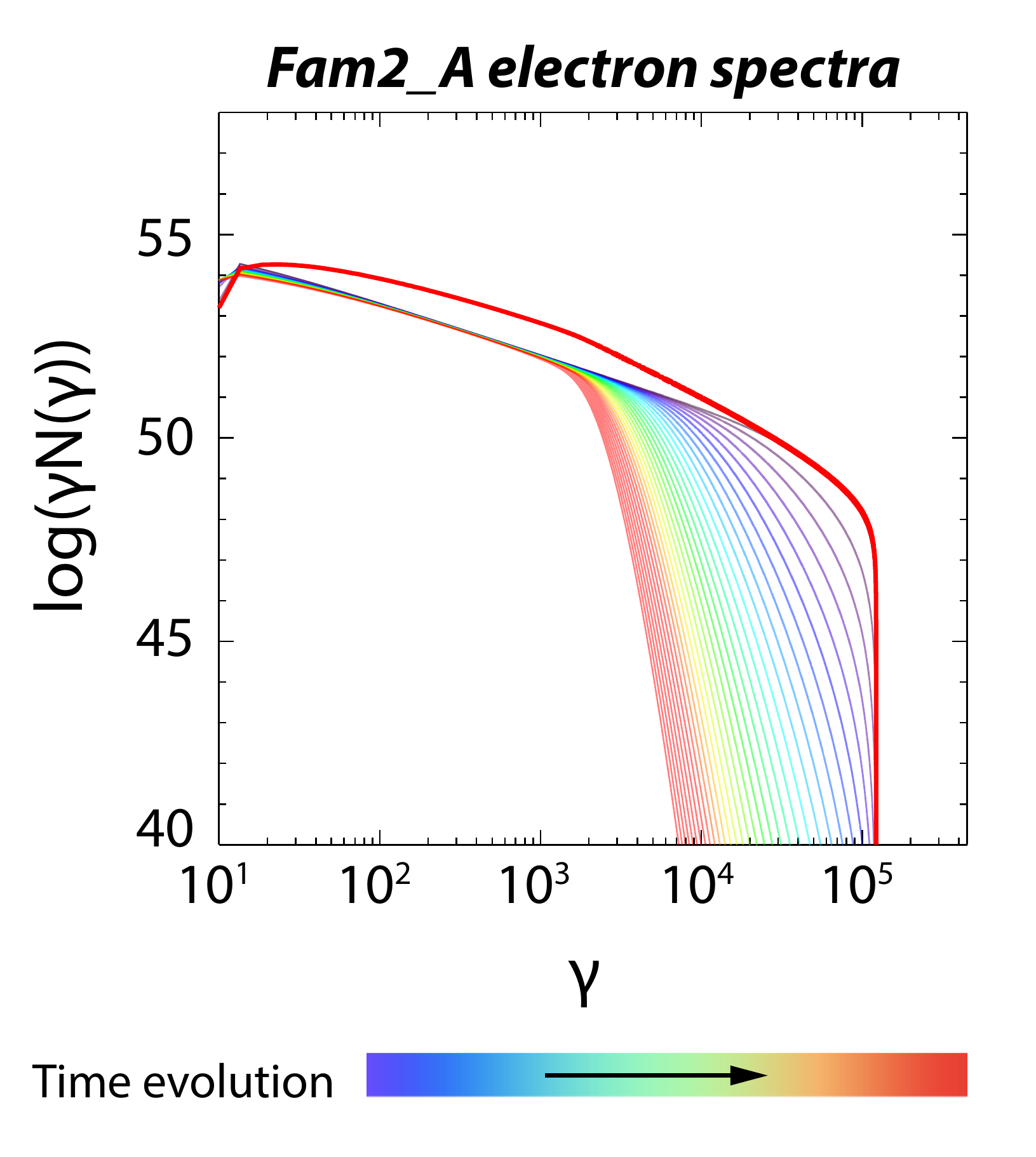}
    \caption{Electron energy spectra time evolution for population-averaged quantities Family 2 tracers in Relic A. Dashed lines correspond to spectra evolution after the first shock. The red solid line represents the electrons spectrum after the second shock.}
    \label{fig:ele_A}
\end{figure}

Figure \ref{fig:ele_A} shows the time evolution of the electron energy spectrum for Family 2 population for Relic A. 
The electron population is produced by the first shock at $t_1=12.69$ Gyrs with a power-low spectrum (purple dashed line of the spectrum) and, as time evolves, we observe a cooling of the high-energy tail of the spectrum, that corresponds to a cut-off at $\gamma\sim10^3$ right before the second shock.
After the final shock at $t_{\mathrm{end}}=13.76$ Gyrs, we observe that the electron energy spectrum is no longer a power-law and electrons are accelerated up to $\gamma\sim10^5$ with a soft knee in the slope around $\gamma\sim10^3$, in correspondence of the cut-off energy before the shock. (red solid line of the spectrum).
However, we are cautious about the result in the low energy part of the spectra, considering the limit of the Fokker-Planck code described in Sec. \ref{subsec:fokker}.


\subsection{Relic B}
For Relic B, the family-averaged quantities are reported in Tab. \ref{tab:relicB_stat}.
We use these quantities to compute the time evolution of the electron energy  spectrum according to the model introduced in Sec. \ref{subsec:fokker}.
\begin{table}
    \centering
    \begin{tabular}{c|c|c|c|c|c}
         & Time [Gyrs] & Mach & B [$\mu$G] & $\rho$ [g/cm$^{-3}$] & T [K]\\
    \hline
    Family 1 & & & & \\
    Shock 1 & 13.76 & 2.7 & $1.6\times10^{-1}$ & $3.1\times10^{-28}$ & $6.5\times10^{7}$ \\
    \hline
    Family 2 & & & & \\
    Shock 1 & 12.82 & 3.5 & $1.9\times10^{-1}$ & $5.1\times10^{-28}$ & $3.2\times10^{7}$ \\
    Shock 2 & 13.76 & 2.8 & $1.4\times10^{-1}$ & $2.7\times10^{-28}$ & $6.6\times10^{7}$ \\
    \hline
    Family 3 & & & & \\
    Shock 1 & 12.56 & 2.4 & $1.4\times10^{-1}$ & $2.8\times10^{-28}$ & $1.9\times10^{7}$ \\
    Shock 2 & 13.31 & 2.4 & $3.8\times10^{-1}$ & $3.7\times10^{-28}$ & $4.0\times10^{7}$ \\
    Shock 3 & 13.76 & 2.8 & $1.5\times10^{-1}$ & $2.9\times10^{-28}$ & $6.7\times10^{7}$ \\
    \hline
    Family 4 & & & & \\
    Shock 1 & 7.82 & 2.3 & $6.2\times10^{-1}$ & $4.9\times10^{-28}$ & $1.0\times10^{7}$ \\
    Shock 2 & 10.98 & 2.9 & $2.0\times10^{-1}$ & $9.4\times10^{-28}$ & $6.4\times10^{7}$ \\
    Shock 3 & 13.37 & 2.1 & $5.6\times10^{-1}$ & $6.2\times10^{-28}$ & $5.5\times10^{7}$ \\
    Shock 4 & 13.76 & 2.3 & $1.8\times10^{-1}$ & $3.4\times10^{-28}$ & $7.5\times10^{7}$ 
    \end{tabular}
    \caption{Family-averaged quantities at selected shock times for families in Relic B.}
    \label{tab:relicB_stat}
\end{table}

\begin{figure}
    \centering
    \includegraphics[width=0.75\columnwidth]{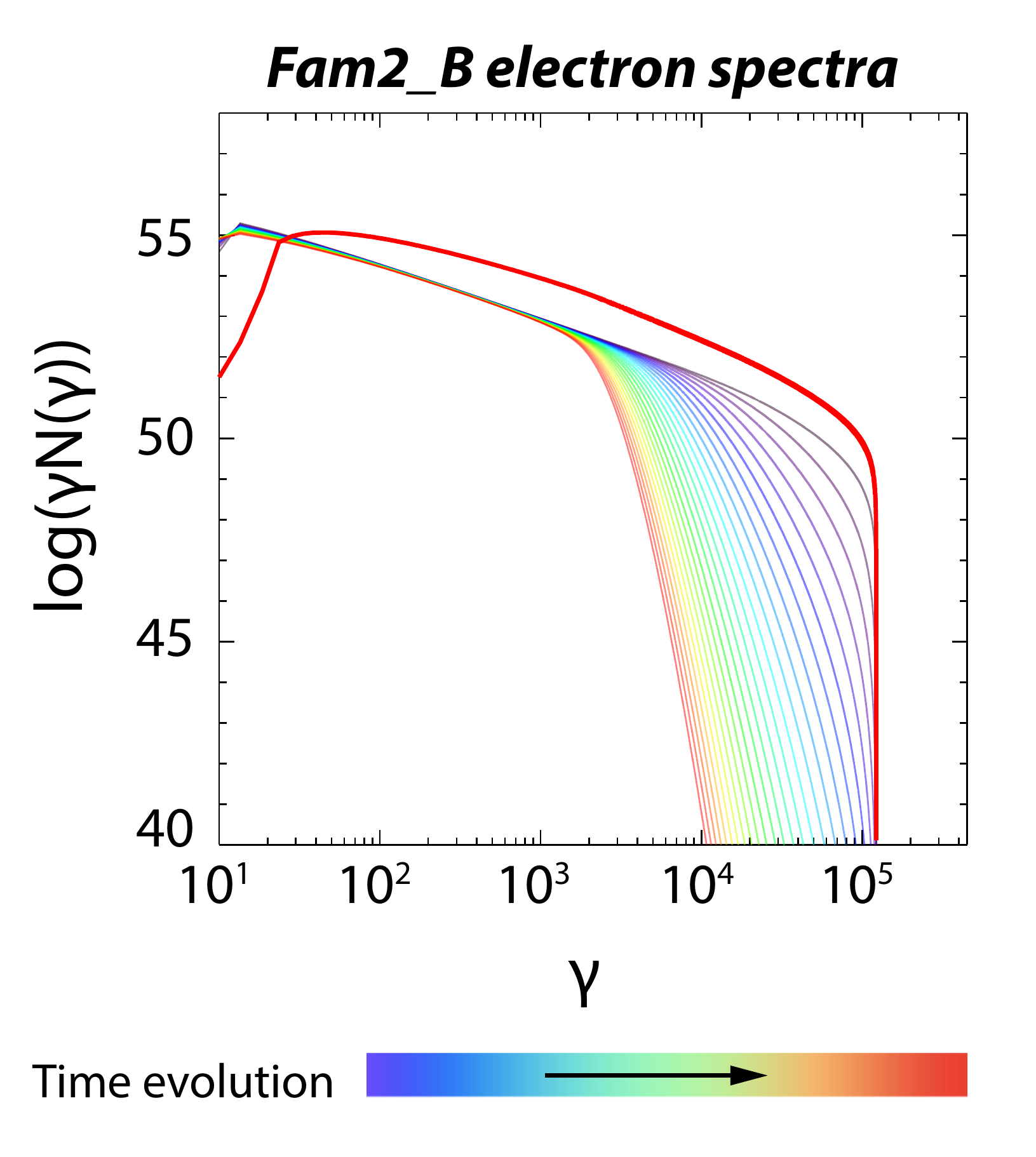}
    \caption{Electron spectra time evolution for quantities-averaged Family 2 tracers for Relic B. Dashed lines correspond to spectra evolution after the first shock. The red solid line represents the electrons spectrum after the second shock.}
    \label{fig:ele_B}
\end{figure}

Figure \ref{fig:ele_B} shows the time evolution of the electron spectrum obtained from the Fokker-Planck model described in Sec. \ref{sec:method} using the averaged quantities for Family 2 tracer population for Relic B as in Tab. \ref{tab:relicB_stat}.

Figure \ref{fig:ele_B} shows the time evolution of the electron energy spectrum for Family 2 population for Relic B. 
We observe a similar behaviour of the electron energy spectrum evolution as for Relic A.
However, since Family 2 population in Relic B is more than one order of magnitude higher than Family 2 population in Relic A, we notice that the absolute value of the electron energy spectrum for Relic B is approximately one order of magnitude higher than Reilc A spectrum.

Similar electron energy spectrum have been obtained for the other Families in Relic B (not shown) in which we observed a behaviour for MS scenario acceleration in the evolution of the spectra consistent with the evolution shown here for Family 2 population.


The family-averaged analysis shown here allowed us to witness the different evolution of MS scenario electron energy spectra compared to a single shock scenario.
However, we noticed that the averaged analysis introduced an huge variation in the computation of the electron energy spectra and, subsequently, in the radio emission associated (see Appendix \ref{sec:appendix}).
In the next Section (Sec. \ref{sec:integrated_radio}) we shall instead compute the detailed radio emission based on the specific sequence of physical fields recorded by each tracer during its evolution, and  compute the integrated radio emission across the relic by combining the information of all tracers in all families.


\section{Integrated radio emission}
\label{sec:integrated_radio}

In this section, we study the integrated radio emission along the same viewing angle of Fig.~\ref{fig:video}, obtained using the electron spectra produced via Fokker-Planck integration over $\sim7000$ tracers (Sec. \ref{sec:method}).
Contrarily to the family-averaged analysis discussed in the previous section, we now compute the energy spectra using the values of density, magnetic field and density recorded by each tracer.
At the final position, we compute the radio emission for both Relic A and Relic B.

\begin{figure*}
    \centering
    \includegraphics[width=\textwidth]{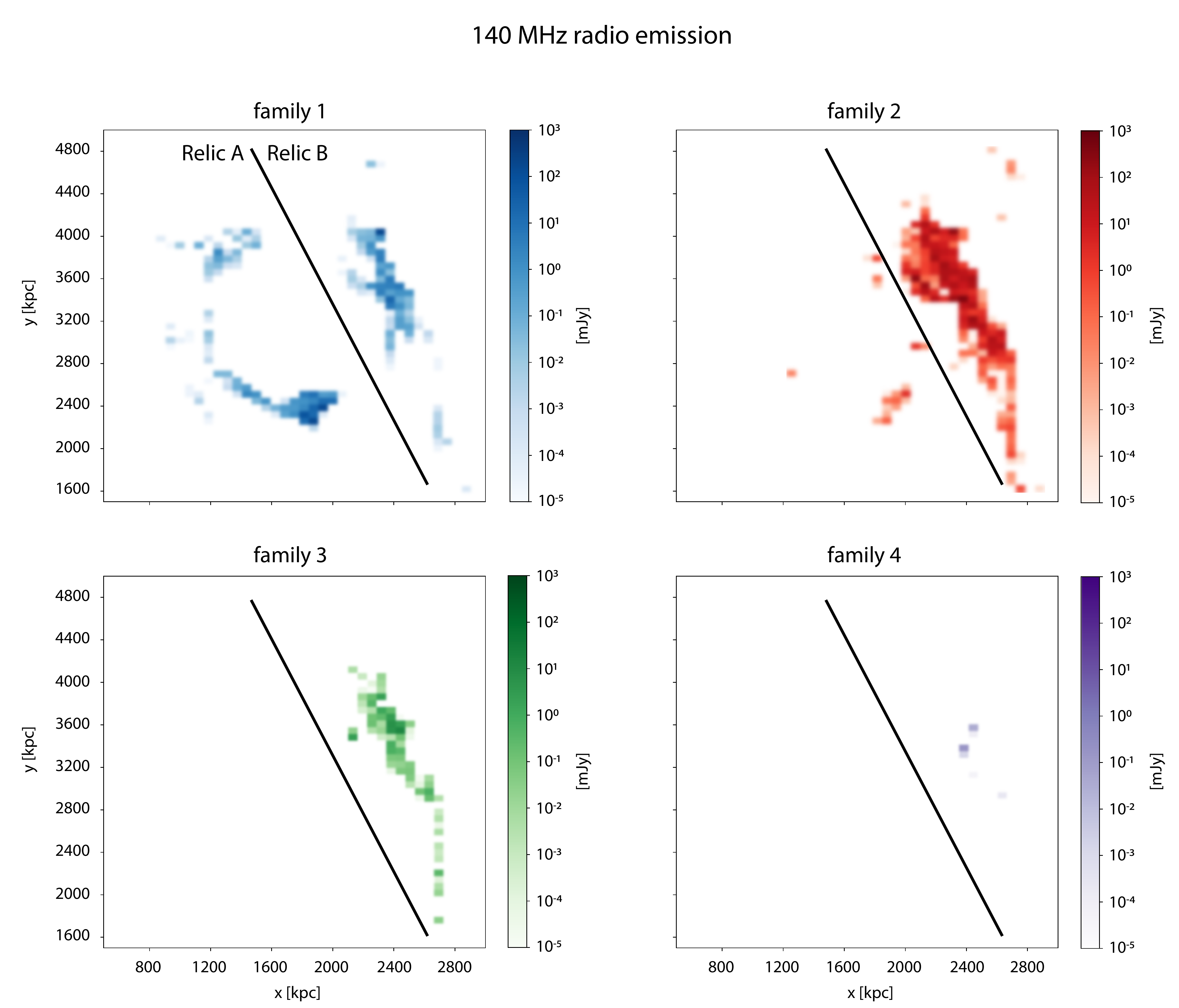}
    \caption{Integrated radio emission $(x,y)$ projection at $140$ MHz for Family 1 (top-left), Family 2 (top-right), family 3 (bottom-left), and Family 4 (bottom-right). The Black solid line in the plots divides the particles population between Relic A and Relic B.}
    \label{fig:radio140}
\end{figure*}

\begin{figure*}
    \includegraphics[width=\textwidth]{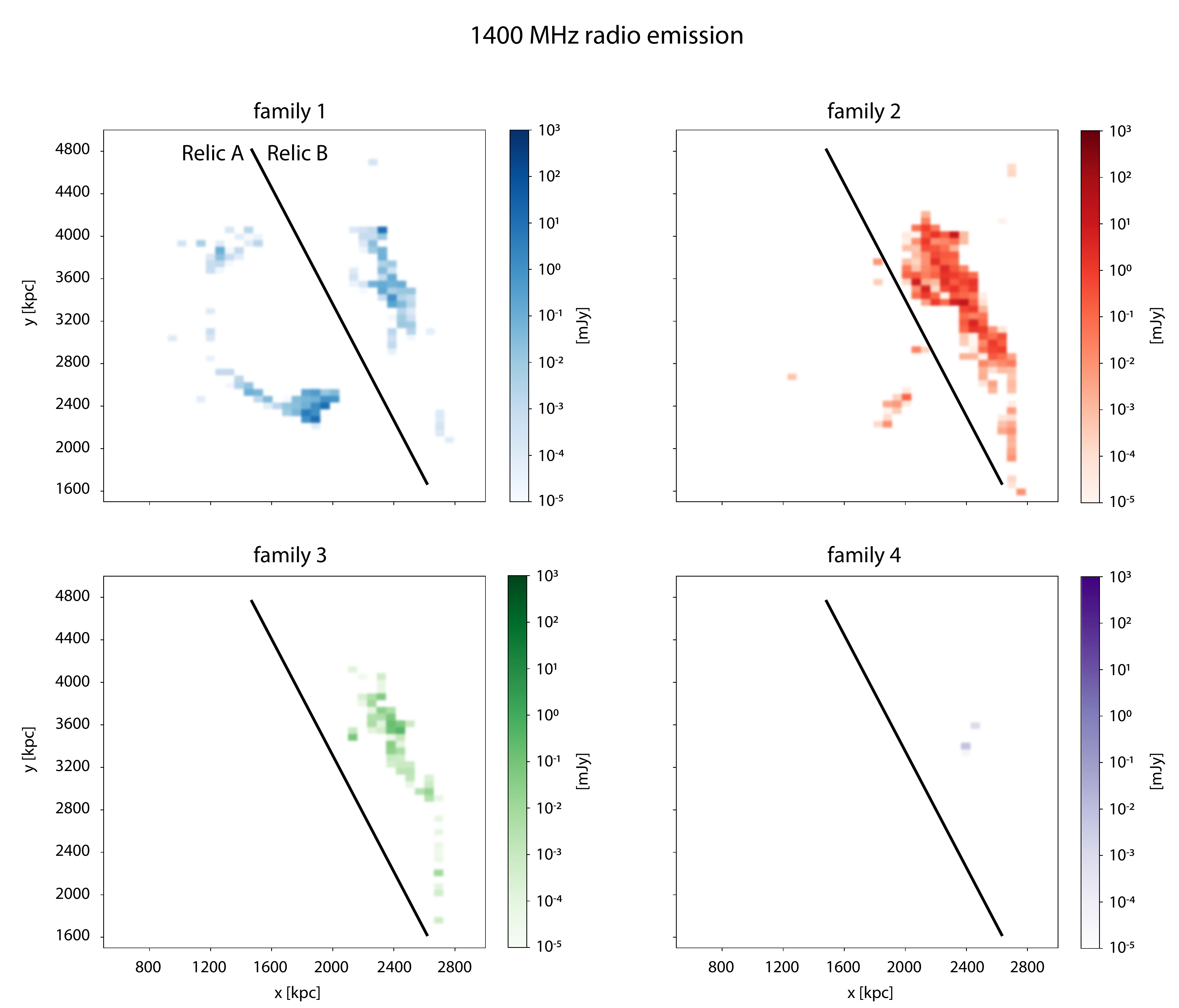}
    \caption{Integrated radio emission $(x,y)$ projection at $1.4$ GHz for Family 1 (top-left), Family 2 (top-right), family 3 (bottom-left), and Family 4 (bottom-right). The Black solid line in the plots divides the particles population between Relic A and Relic B.}
    \label{fig:radio1400}
\end{figure*}

Figures \ref{fig:radio140} and \ref{fig:radio1400} show the integrated radio emission maps, respectively, at $140$ MHz and $1.4$ GHz, in which we separate the emission contributions from the different families. 
Assuming the distance of the observation at $z=0.15$, we calculate the integrated radio emission with a beam size of $63.7$ kpc, corresponding to the $25"$ LOFAR telescope resolution.

Focusing on Relic A, we see that the integrated radio emission is mostly dominated by Family 1 population and reaches a peak of $\lesssim 10^3$ mJy at $140$ MHz, while radio emission of Family 2 is concentrated in the lower-right corner of the relic, and its integrated value is $\sim$one order of magnitude lower.  This makes us conclude that, in Relic A, the visible radio emission will be mostly dominated by Family 1 population, i.e. by newly accelerated electrons.
This object appears therefore as a ``classic" powerful radio relic, in which all/most of the observed emission is due to the latest shock, which has energised a pool of fresh electrons, which are being observed within a cooling time since their first acceleration. 

Interestingly, the situation is very different for the nearby  Relic B, whose integrated radio emission of $\sim 10^3$ mJy at $140$ MHz is dominated by the Family 2 population.
The electrons from Family 1 and Family 3 are confined in a small sub-volume of Relic B, albeit they have an overall comparable radio emission between each others.
The emission for Family 4, instead, due to its smaller occupation fraction,  remains negligible at all frequencies. 

\begin{figure}
    \centering
    \includegraphics[width=\columnwidth]{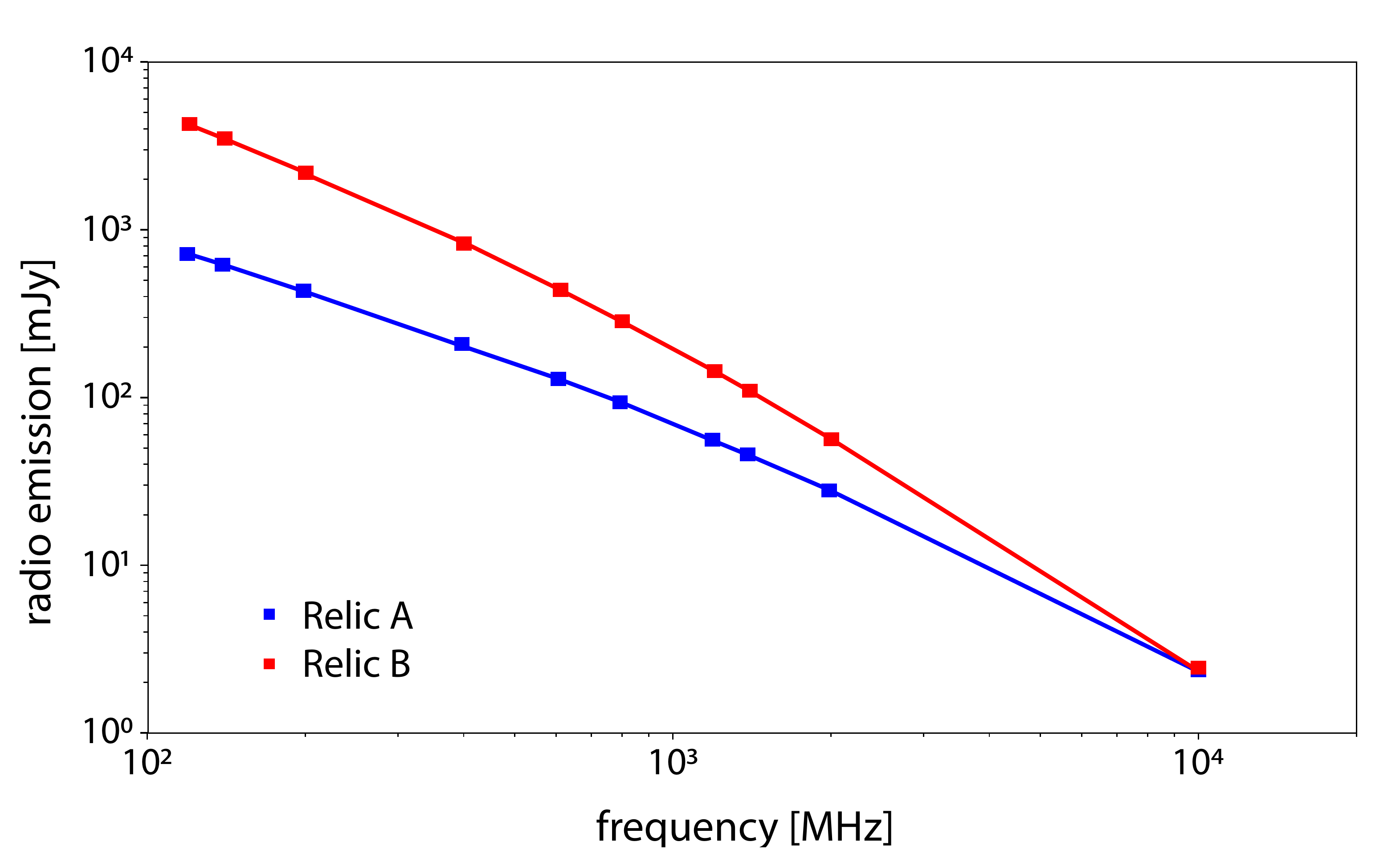}
    \caption{Relic-integrated radio spectrum for Relic A (blue) and Relic B (red).}
    \label{fig:tot_spectra}
\end{figure}
In Figure \ref{fig:tot_spectra}, we give the total emission for the ``single" zone analysis of the two relics, i.e. by integrating the CRe emission over the volume of each relic.
In a single zone and standard view of radio relics, a $\langle \alpha \rangle \sim -1.55$ associated with a shock with $\mathcal{M} \approx 2.1$.
In order for the shock to produce the observed emission of 96 mJy at 1.4 GHz, a single zone \citet{hb07} method requires to dissipate a fraction in the $K_{inj,e} \approx 3\times10^{-4}-10^{-3}$ ballpark of the kinetic energy flux across the shock into electron acceleration - which is very large for such a weak shock, based on DSA \citep[e.g.][]{2020JKAS...53...59K}.
This is a common finding of real observations, which have also routinely reported requirements on the acceleration efficiency even of $\sim 100 \%$, or larger, in several objects \citep[e.g.][]{2016MNRAS.461.1302E,Stuardi2019,2020A&A...634A..64B}.
Our analysis instead shows that, even in absence of a nearby active galactic source of radio electrons, sectors of galaxy clusters interested by the MS scenario crossing can boost the emission to a level compatible with observations, due to the re-acceleration of fossil particles injected at a $\leq 0.5-1 \rm ~Gyr$ time interval.
Detailed values of the single zone radio emission for the two relics at indicative frequencies are reported in Tab. \ref{tab:relic_emission}.

\begin{table}
    \centering
    \begin{tabular}{c|c|c}
         & Relic A & Relic B \\
         \hline
        140 MHz & 515 mJy & 2768 mJy \\
        400 MHz & 179 mJy & 684 mJy \\
        1.4 GHz & 41 mJy & 96 mJy 
    \end{tabular}
    \caption{relics-integrated radio emission.}
    \label{tab:relic_emission}
\end{table}

\bigskip
\begin{figure*}
    \centering
    \includegraphics[width=0.49\textwidth]{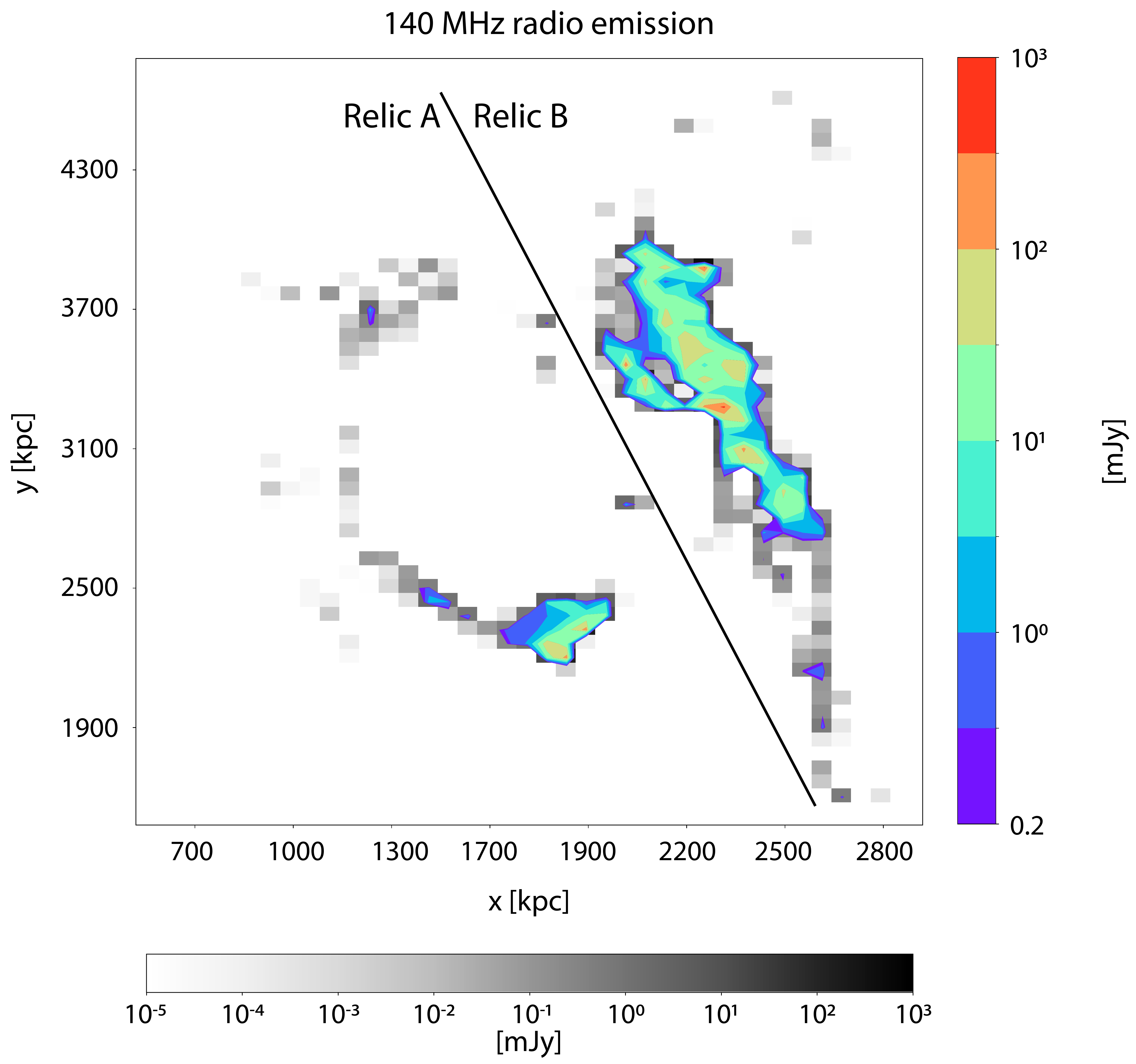}
    \includegraphics[width=0.49\textwidth]{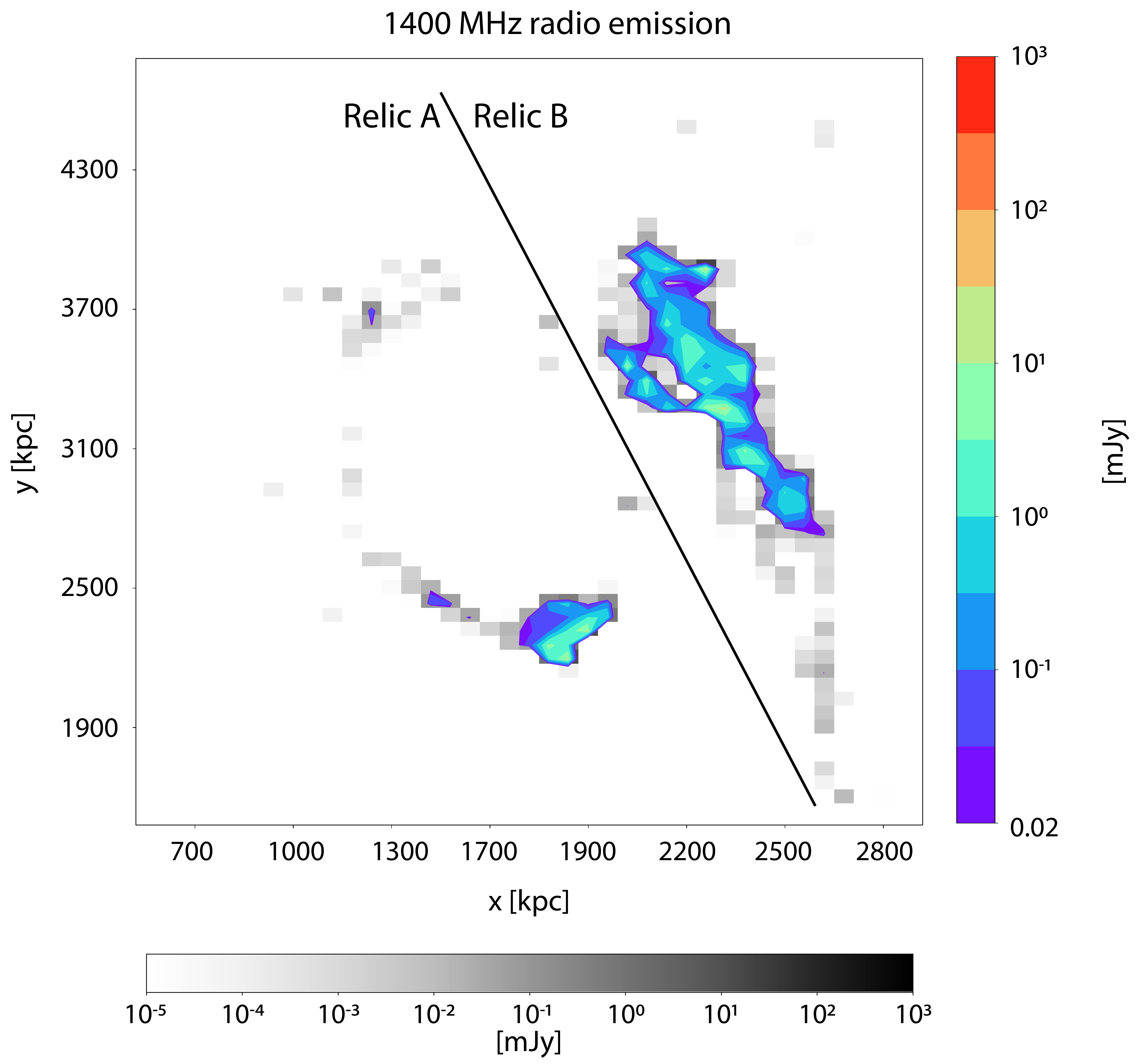}
    \caption{Integrated radio emission for all families at $140$ MHz (left) and $1.4$ GHz (right) in gray scale. The coloured contours in the two plots indicate the observable radio flux using the LOFAR threshold of $0.2$ mJy at $140$ MHz and $0.02$ mJy at $1.4$ GHz respectively. The Black solid line in the plots divides the particles population between Relic A and Relic B.}
    \label{fig:emission_observed}
\end{figure*}
We investigate the possibility to observe the simulated relics by comparing the integrated radio emission obtained from the Fokker-Planck model with LOFAR observation properties.
Figure \ref{fig:emission_observed} shows the integrated radio emission with the contribution of all the families in the gray scale.
On top of that, the coloured contours indicate the observable radio emission using the LOFAR threshold of $0.2$ mJy at $140$ MHz and JVLA threshold of $0.02$ mJy at $1.4$ GHz respectively.
We can conclude that part of the two radio relics generated in our numerical studies are bright enough to be observable.

\subsection{Spectral Index Map}
To investigate possible differences in the spectral index properties at the shock and the energy losses in the post-shock region, we analyzed the spectral index profile across the radio relics.
We obtain the spectral index map by fitting with a first-order polynomial (i.e. $y=a_1x + a_0$) the integrated radio emission calculated in Sec. \ref{sec:integrated_radio} between frequencies $140$\,MHz and $400$\,MHz, and between $400$\,MHz and $1.4$\,GHz.
For a first-order polynomial fit, the spectral index $\alpha$ is defined as the slope of the fit, i.e., $\alpha=a_1$.
Using this first order fit, we obtain that the radio emission $I$ can be calculated locally at each frequency $\nu$ as $I\propto\nu^{\alpha_1}$.

\begin{figure*}
    \centering
    \includegraphics[width=0.49\textwidth]{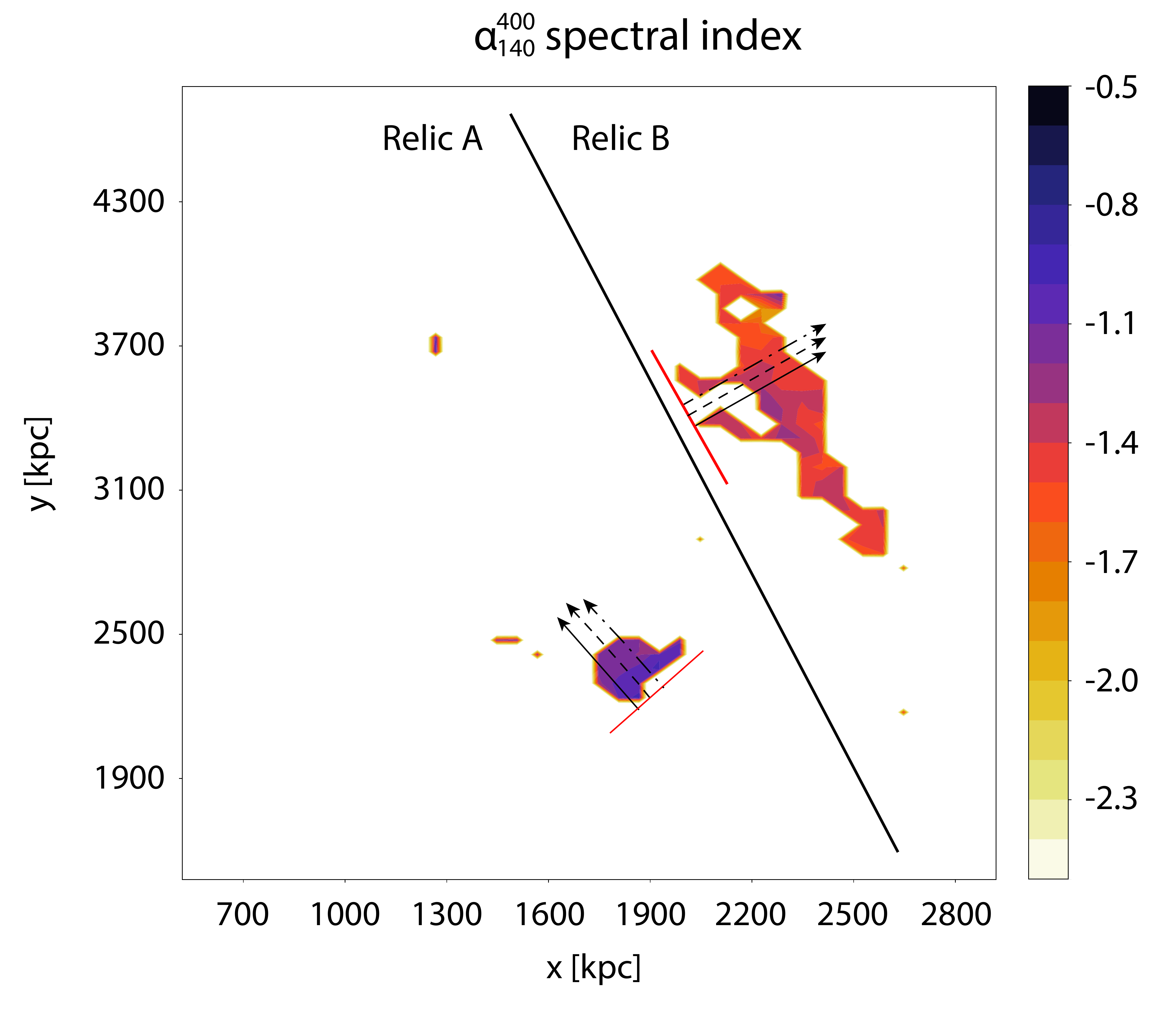}
    \includegraphics[width=0.49\textwidth]{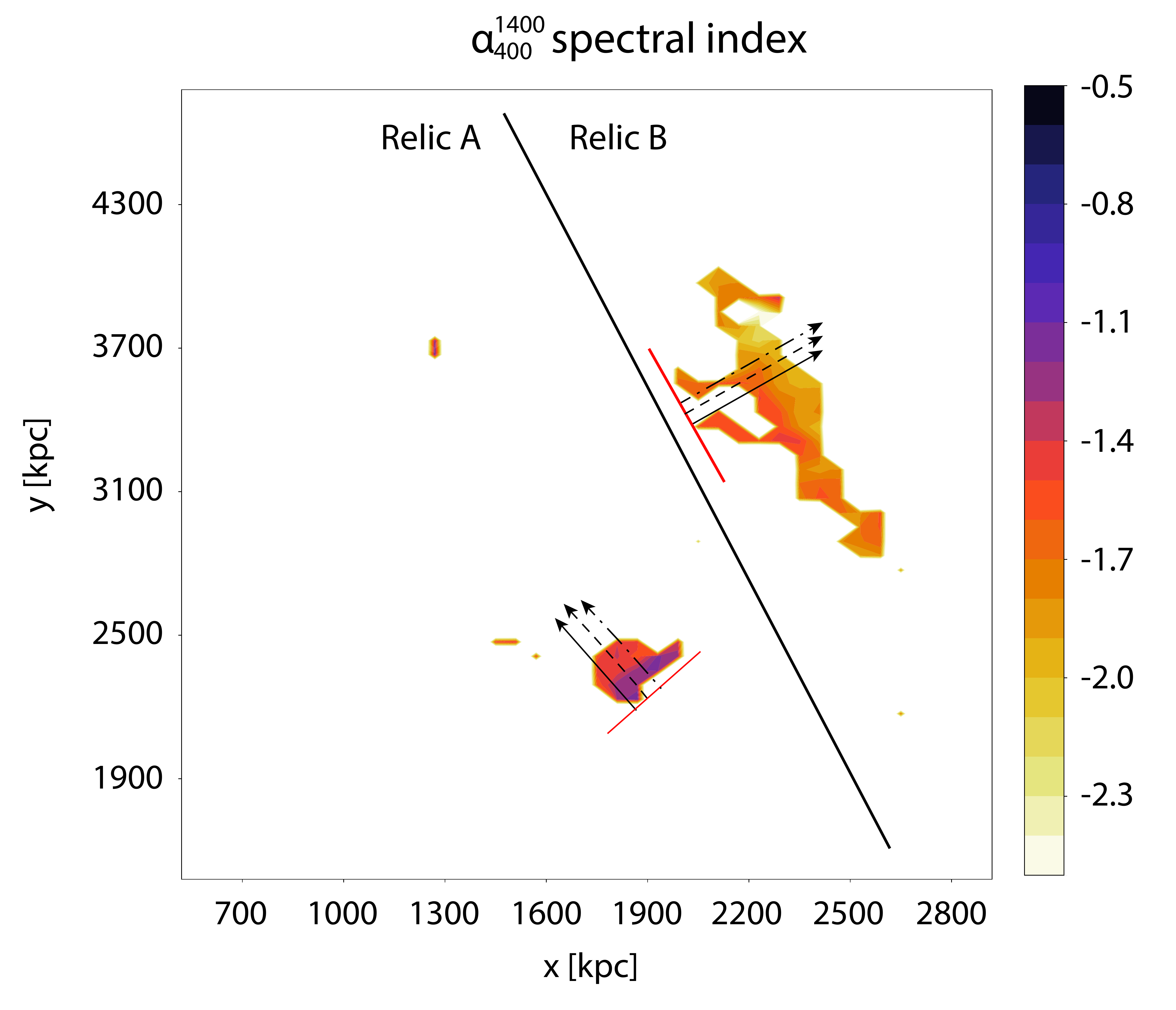}
    \caption{(left) $\alpha_{140\,\rm MHz}^{400\,\rm MHz}$ spectral index map and (right) $\alpha_{400\,\rm MHz}^{1.4\,\rm GHz}$ spectral index map from the contribution of all families population. The red lines indicate the position of the shock front for each relic, while} the black lines over the relics indicate the lineout of Fig. \ref{fig:index_profile}.
    \label{fig:spectral_index}
\end{figure*}
Figure \ref{fig:spectral_index} shows the $\alpha_{140\,\rm MHz}^{400\,\rm MHz}$ and $\alpha_{400\,\rm MHz}^{1.4\,\rm GHz}$ spectral index maps,  obtained from the contribution of all families populations.

The two relics have rather distinct  spectral index properties, which may reflect the different histories of shock acceleration in the two cases.
Relic A, dominated by family 1 population, shows a spectral index in the range  $-1.0$ to $-1.2$ between  $140$ and 400 MHz and  $-1.2$ to $-1.5$ between 400 MHz and $1.4$ GHz.
Relic B, dominated by family 2 population instead, shows a spectral index in the range of $-1.2$ to $-1.5$ between  140 and 400 MHz and $-1.4$ to $-2.0$ between 400 MHz and 1.4 GHz. In particular, the presence of multiple shocks acceleration events makes Relic B brighter than Relic A, despite having a steeper radio spectral index. 

\begin{figure*}
    \centering
    \includegraphics[width=0.49\textwidth]{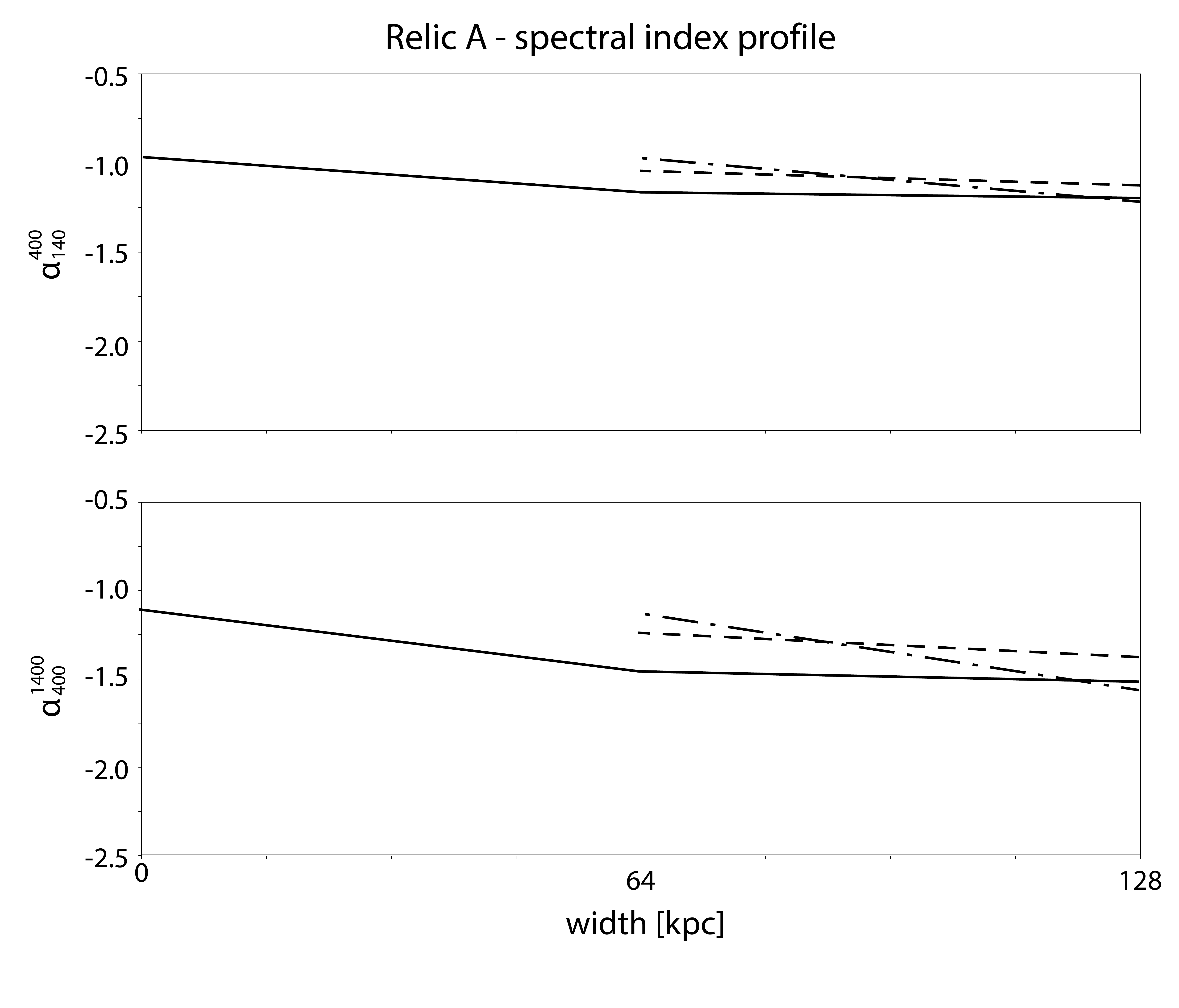}
    \includegraphics[width=0.49\textwidth]{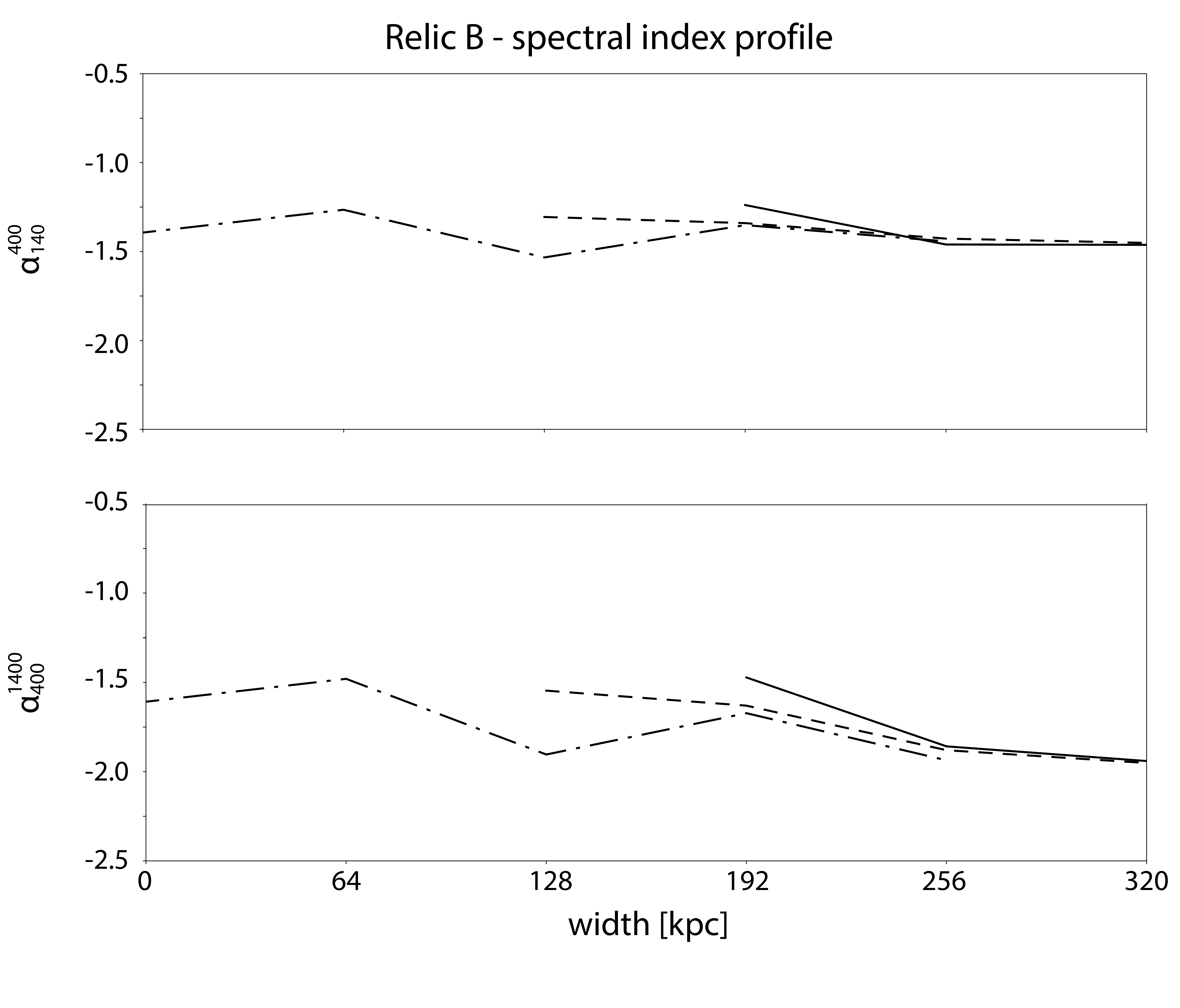}
    \vspace{-.6cm}
    \caption{Spectral index profile for Relic A (left) and Relic B (right) between two different frequency ranges. The different black lines correspond to the profile position for each relic as indicated in Fig. \ref{fig:spectral_index}. The origin of the profiles correspond to the shock front location for each relic.}
    \label{fig:index_profile}
\end{figure*}
Figure \ref{fig:index_profile} shows the spectral index profile for Relic A and Relic B at the line-outs indicated by black lines in Fig. \ref{fig:spectral_index}.
For each relic, we indicated the position of the shock front and its propagation with a red line in Fig. \ref{fig:spectral_index} and we compute the line profiles along the direction perpendicular to the shock front, starting at its position.
For Relic A, we observe an almost constant distribution of the spectral index along its profile around a value of $-1$ at low frequency, and at $\sim -1.2$ at high frequency, however the resolution is not high enough for this relic to make a strong statement from such finding.
For Relic B, instead, we notice that the spectral index has a generally more fluctuating behaviour,  with a steeper radio spectrum at both frequencies, in particular between $\sim -1.3-1.4$ at low frequency and between $\sim -1.5-2.0$ at high frequency, which remains nearly constant even $\geq 200$ $\rm kpc$ away from the shock edge. The latter behaviour appears as a natural consequence of the MS scenario, in the sense that the emission of Relic B is dominated by the low-energy component of re-accelerated electrons, whose radio emission remain high also away from the shock edge. However, since the time elapsed since the epoch of the first injection of electrons in the MS scenario can vary from case to case, depending on the specific accretion history of the host cluster, and on the dynamics in the cluster sector where relics form, different timings of accretions should be reflected in different steepening frequencies for real observed relics.

\subsection{Radio colour-colour Diagram}
The shape of the relativistic electron distribution in radio sources can be  studied by means of so-called colour–colour diagrams \citep{Katz1993, Rudnick1996, Rudnick2001}.
These diagrams emphasize the spectral curvature because they represent a comparison between spectral indices calculated at low- and high-frequency ranges. 
\begin{figure*}
    \centering
    \includegraphics[width=0.49\textwidth]{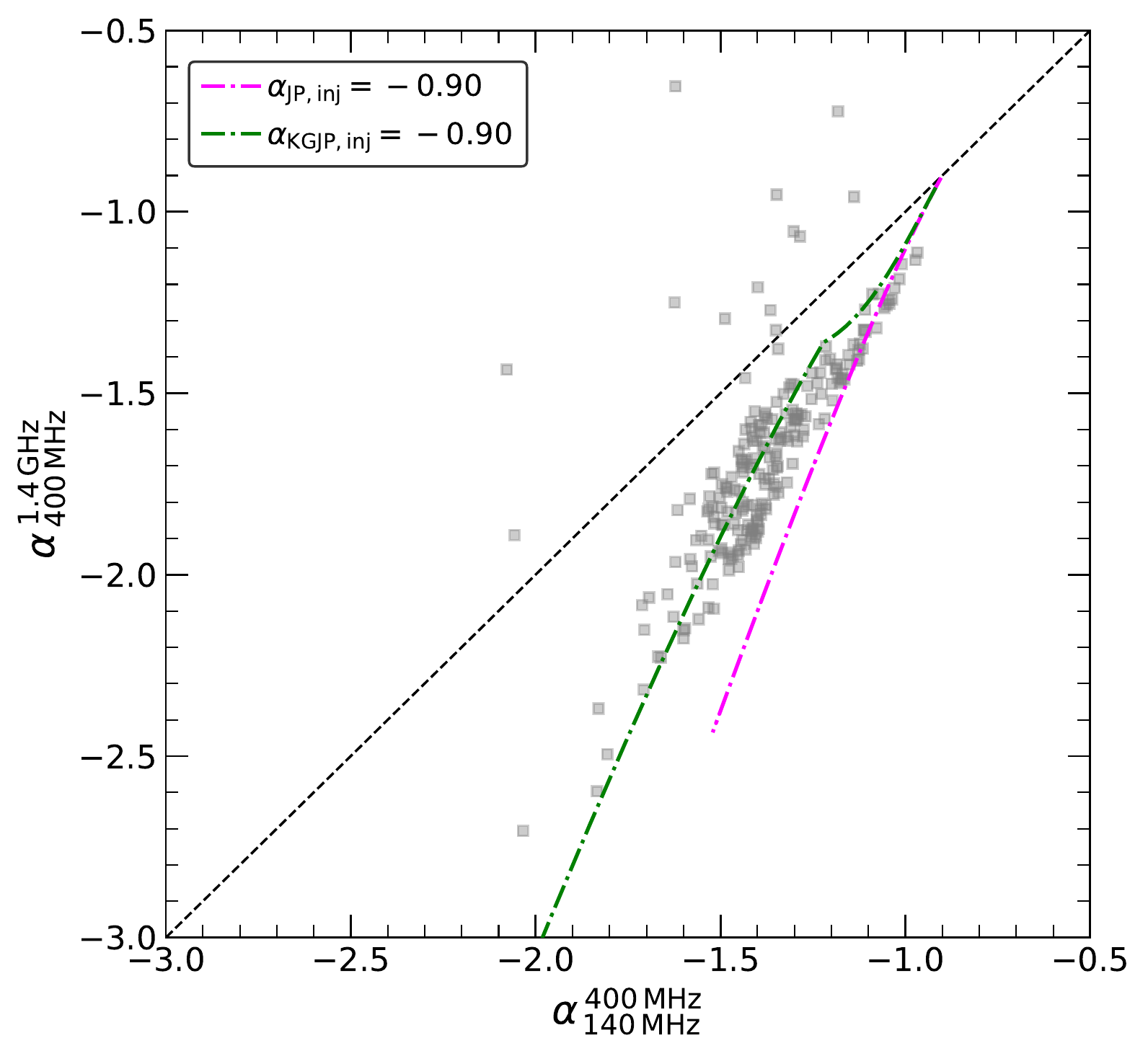}
    \includegraphics[width=0.49\textwidth]{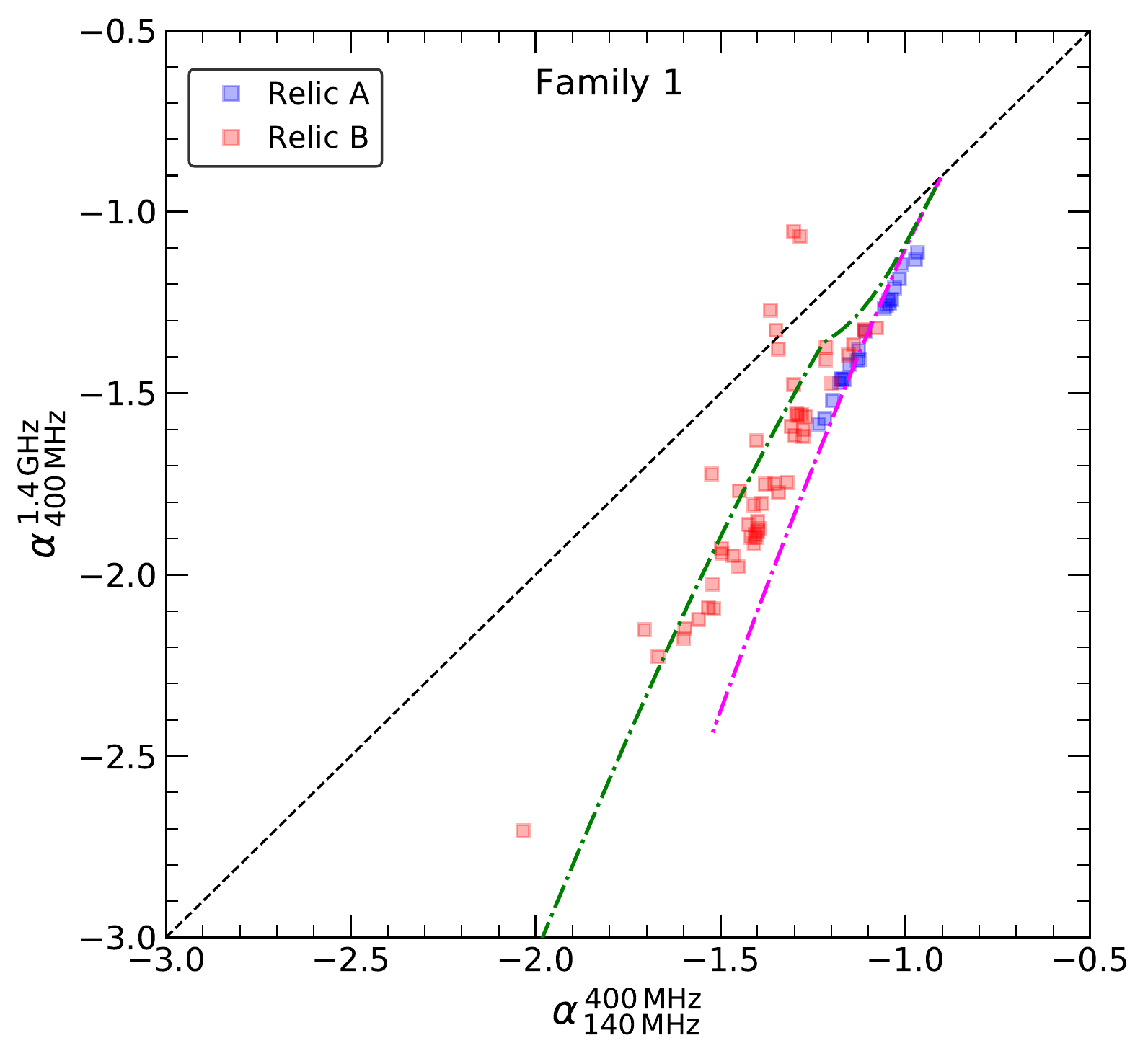}
    \includegraphics[width=0.49\textwidth]{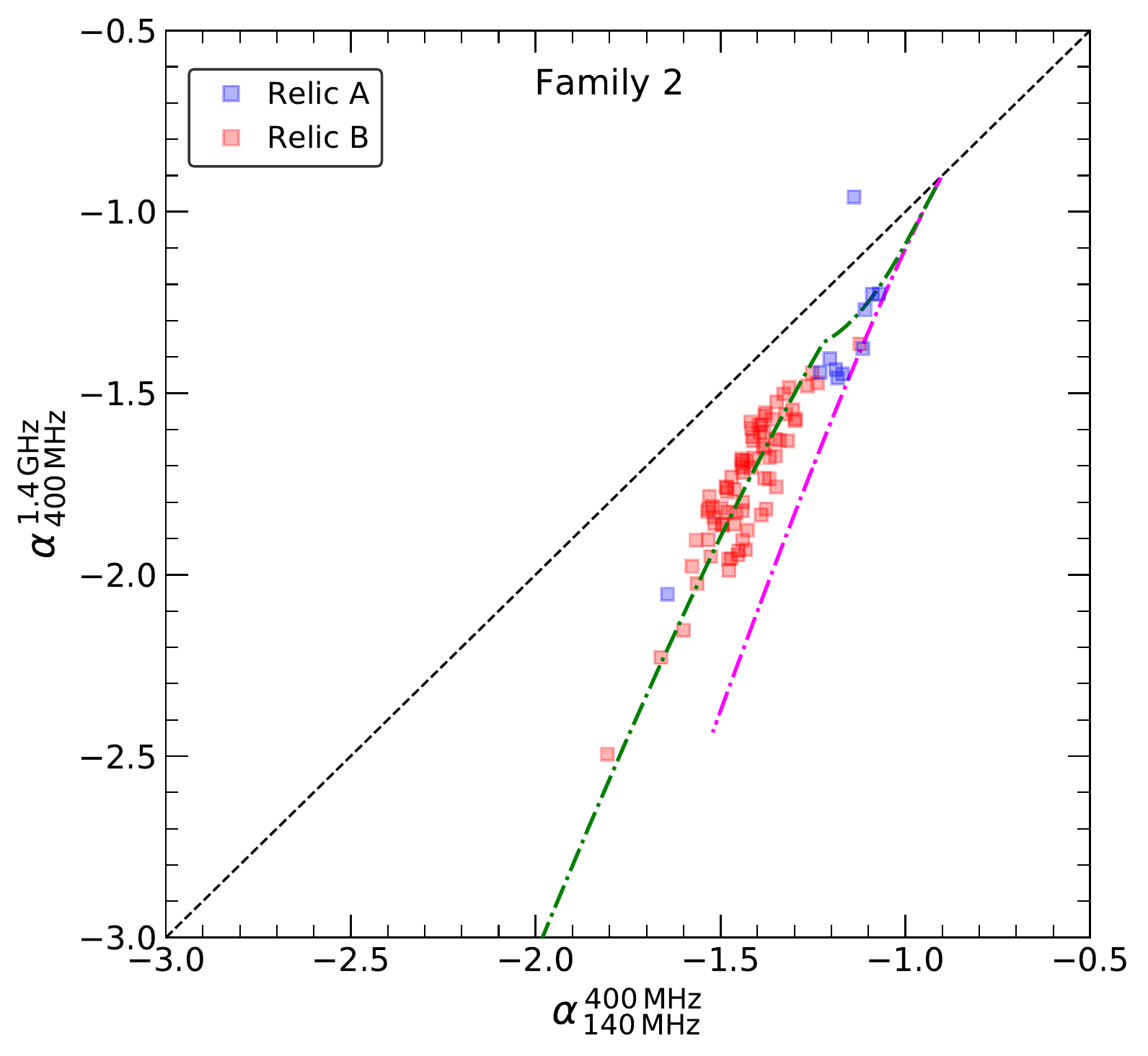}
    \includegraphics[width=0.49\textwidth]{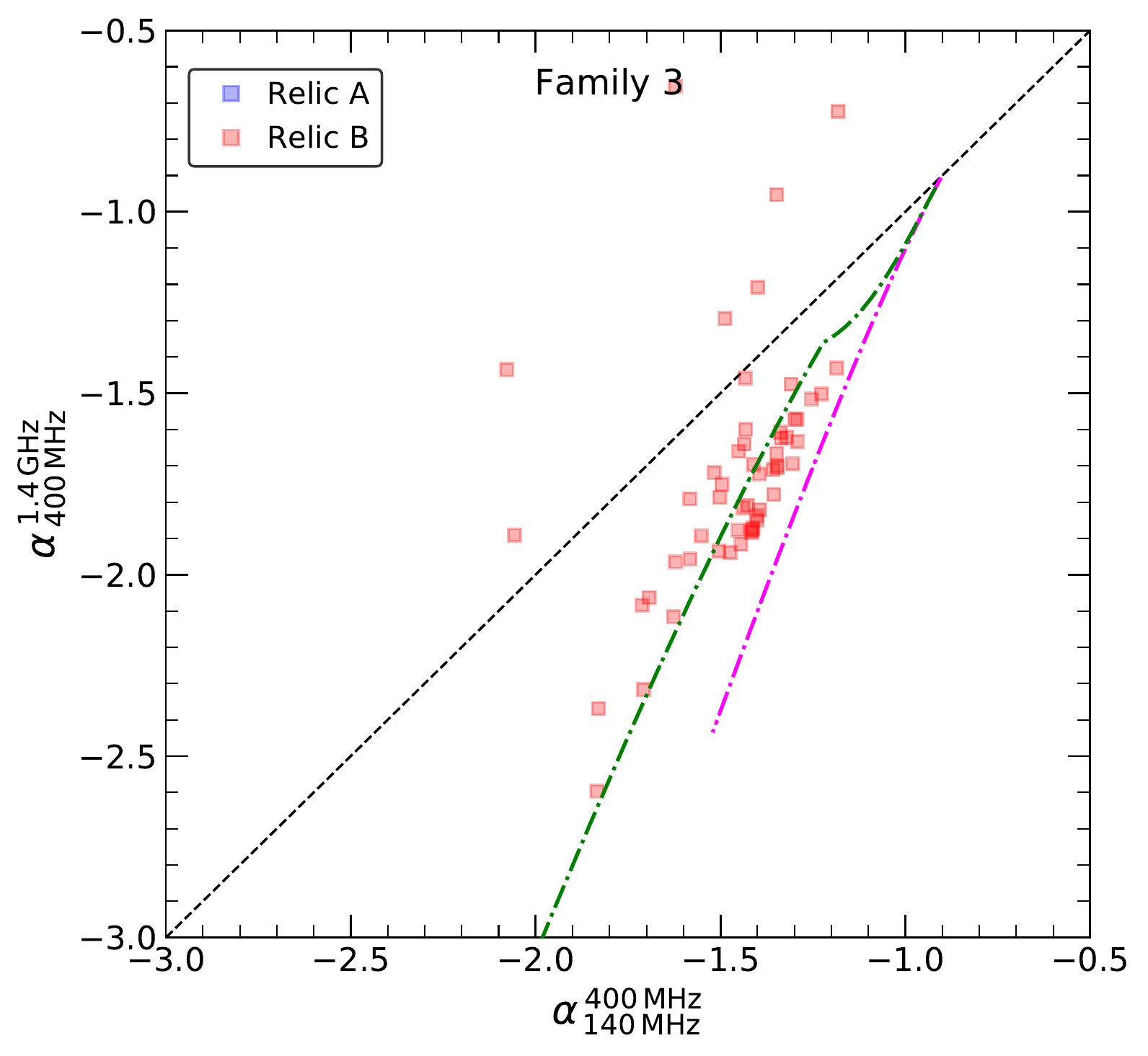}
    \caption{Radio colour-colour plots of the simulated relic superimposed with the spectral ageing JP (magenta) and KGJP (green) models obtained using an injection index of $-0.90$. For KGJP model, particles are injected continuously for about 16\,Myr.  Colour-colour plot for all families (top-left), Family 1 (top-right), Family 2 (bottom-left), and Family 3 (bottom-right). The KGJP model fits the distribution well.}
    \label{fig:color2}
\end{figure*}
In our case, the low frequency spectral index values were calculated between 140 and 400\,MHz while the high frequency one between 400\,MHz and 1.4\,GHz.
By this convention, the curvature is negative for a convex spectrum.
The resulting colour-colour plots are shown in Fig. \ref{fig:color2}.
The dashed black line indicates a power-law spectrum where  $\alpha_{140\,\rm MHz}^{400\,\rm MHz}=\alpha_{400\,\rm MHz}^{1.4\,\rm GHz}$.
Any curve deviating from the power-law line represents a spectrum with changing spectral curvature. 

As visible in Fig.\,\ref{fig:color2} (top-left), we find a clear negative curvature as also reported for some of the well-known relics, for example the Toothbrush \citep{rajpurohit2020toothbrush}, the Sausage \citep{digennaro2018saus}, and MACS\,J0717.5+3745 \citep{rajpurohit2021macsspec}. The single continuous trajectory in the colour-colour plot suggests that the spectrum also has single shape.
We also superimposed the resulting plot with the conventional spectral ageing models, namely JP and KGJP \citep{Komissarov1994}, adopting an injection index of $-0.90$.
The JP model assumes a single burst of particle acceleration and a continued isotropization of the angle between the magnetic field and the electron velocity vectors (the so-called pitch angle) on a timescale shorter than the radiative timescale.
An extension to the JP model is the KGJP that includes finite time of particle injection. 

In Fokker-Planck model (see Section\,\ref{sec:method}) used for the computation of the radio  spectra, we use a JP model for the synchrotron energy losses \citep{1973A&A....26..423J}.
At first sight, it may seem surprising that the JP model does never fit the data (Fig.\,\ref{fig:color2}).
As discussed in \cite{rajpurohit2020toothbrush,rajpurohit2021macsspec}, a perfectly edge-on shock front with a uniform magnetic field can be described by the JP model. However, if the shock front is inclined with respect to the line of sight, different spectral ages are present and contribute to the observed spectrum.
In this case, the colour-colour distribution follows the KGJP model.
As seen in Fig.\,\ref{fig:color2}, the KGJP model with an injection index of $\approx -0.90$ can describe the entire distribution quite well, consistent with what is found for the Toothbrush and the Sausage relics  \citep{rajpurohit2020toothbrush,digennaro2018saus}.  
     
Among the relics, Relic B shows the maximum curvature but  both relics follow the same single curve.
We do not find any significant difference in the curvature distribution between different families, see Fig. \ref{fig:color2}.
We note that there are no data points in the range $-0.5$ to $-0.90$ for both low and high frequency spectral index values.  
This difference can be understood considering that the radio emission properties derived by our Fokker-Planck model have the intrinsic limitation that all particles (even the one injected by the latest shock in the simulation) are evolved at least for one timestep.
Hence, even the youngest population of electrons in both our relics has evolved for  one timestep after shock injection, with a duration $\rm \Delta t \approx 30~ \rm Myr$, and the effects of synchrotron and Inverse Compton losses on the observed radio spectrum are already visible at radio emitting frequencies. 
In summary, the fact that both our relics reasonably well compare with the colour-colour diagrams of real systems (and especially the circumstance that our Relic B is in line with the KGJP model) further confirms that the MS scenario acceleration explored in this work may indeed give rise to realistic relic configurations - albeit non-trivial to tell apart from single injection models, at least based on their colour-colour plots.

\section{Conclusions}
\label{sec:conclusion}
We have simulated the evolution of a radio emitting population of relativistic electrons accelerated by multiple merger shock waves, released during the formation of a  massive, $M_{200} \approx 9.7 \cdot 10^{14} \ \mathrm{M}_{\odot}$, galaxy cluster  \citet[][]{2016Galax...4...71W,2017MNRAS.464.4448W}.
We focused on the spatial and dynamical evolution of $\sim10^4$ tracers, which are located in luminous relic-like structures at the end of our run. 
We assumed DSA as a source of  fresh relativistic electrons out of the thermal pool, and applied a Fokker Planck method to integrate their energy evolution under radiative losses, and further re-acceleration events by merger induced shocks.
In our scenario, only shock waves can be the source of cosmic-ray electrons,  yet multiple shock waves sweeping the ICM may produce a pre-acceleration of relativistic electrons, qualitatively similar to what radio galaxies are expected to do  \citep[e.g.][]{2005ApJ...627..733M, 2011ApJ...728...82M, kr11, ka12, 2014ApJ...788..142K, 2013MNRAS.435.1061P, 2020A&A...634A..64B}.
In particular, we indeed identified a specific multiple-shock (MS) scenario, in which particles cross a shock multiple times, before ending up in realistic $\sim ~ \rm Mpc$-sized radio relics.
Depending on the number of MS events, CRe with a different evolution can become radio visible. This is regardless of the strength of the final shock event. 
One of our relics (Relic A) is found to be mostly dominated by a population of tracers which were  shocked only just before the epoch of the relic formation, and has a very faint emission, only partially detectable with LOFAR. In this respect, this object appears similar to the recently discovered "Cornetto" relic \citep[][]{locatelli2020dsa}, which was suggested to be indeed the prototype of low-surface brightness radio relics, only powered by freshly injected electrons. 

On the other hand, we measured that the emission by a second relic in the system (Relic B) is dominated by MS scenario accelerated electrons. 
We use the shock information collected with the tracers to study the evolution of relativistic electrons injected at the shocks and the associated radio emission via the Fokker-Planck solver described in \ref{subsec:fokker}.
We observe that the electron energy spectrum for MS scenario accelerated families differs significantly from the power-law spectrum obtained after a single shock injection, and that their emission is higher than the emission of electrons that were only shocked at the end of the simulation, up to at least $\sim1$ order of magnitude. 

We computed the total radio emission produced by all accelerated electron families in both relics,  emulating the threshold parameters of LOFAR telescope at $140$ MHz and of the JVLA at $1.4$ GHz, obtaining that both relics can be detected by observations, in particular at lower frequencies.

From the analysis of the spectral index maps, we observed that Relic A shows relatively flat spectral index values compared to Relic B, suggesting that the presence of a MS scenario evolution of fossil electrons may influence the slope of the radio spectrum observed in the different relics.
The radio colour-colour analysis revealed a single continuous curve for both Relics A and B as well as for all families.
The curvature distribution can be well explained  by the KGJP model. 

This suggests that, at least in systems whose past evolution is characterised by multiple accretion events, for example objects with prominent filamentary accretions and a past with multiple mergers, such as Abell 2744 \citep{2004MNRAS.349..385K,raj21}, Coma \citep{2011MNRAS.412....2B,bo21}, the Toothbrush cluster \citep{2012A&A...546A.124V,rajpurohit2020toothbrush}, a significant fraction of the observed radio emission can be the product of MS scenario acceleration, with the effect of an apparent boost of the acceleration efficiency  with respect to ``single shock" models. 
Even if this work is exclusive of a single simulation, we found differences between the radio spectra produced in the MS scenario and the single shock scenario. In particular in the MS scenario, electrons produce a more emission at low frequencies and, hence, a steeper spectrum than the single shock scenario.
This is particularly intriguing, as, in the MS scenario, the radio emission is produced without including any other sources of fossil electrons in the ICM. This can soften the assumption that single radio galaxies are the source of fossil electrons, for the observed cases that require a high acceleration efficiency. The MS scenario can indeed produce a large pool of pre-accelerated electrons, with rather similar spectra and energy densities on $\sim \rm ~Mpc$ scales, which further produce coherent radio properties on the same scales, if further shocked.
The latter may instead be a problem for models in which the source of fossil electrons is a single and recent release of fossil electrons from a radio galaxy.
In reality radio galaxies do exist and inject fossil electrons, and we  defer the investigation of the MS scenario combined with radio galaxy activity to future work.

\section*{Acknowledgements}

We gratefully acknowledge the very useful feedback by our reviewer, H. Kang, which significantly improved our numerical analysis since our first submitted version.
We acknowledge financial  support by the European Union’s Horizon 2020 program under the ERC Starting Grant ‘MAGCOW’, no. 714196.
D.W. is funded by the Deutsche Forschungsgemeinschaft (DFG, German Research Foundation) - 441694982. 
The cosmological simulations were performed with the ENZO code (\hyperlink{http://enzo-project.org}{http://enzo-project.org}) and were partially produced at Piz Daint (ETHZ-CSCS, \hyperlink{http://www.cscs.ch}{http://www.cscs.ch}) in the Chronos project ID ch2 and s585, and on the JURECA supercomputer at the NIC of the Forschungszentrum Jülich, under allocations nos. 7006, 9016, and 9059.
For the creation of the NRAO video, GI acknowledge the assistance of $(AM)^2$ research group of Prof. S. Morigi of the Department of Mathematics of the University of Bologna and the Visit Lab of A. Guidazzoli of CINECA under the project FIBER OF THE UNIVERSE (\hyperlink{http://visitlab.cineca.it/index.php/portfolio/fiber-of-the-universe/}{http://visitlab.cineca.it/index.php/portfolio/fiber-of-the-universe/}). We acknowledge the usage of online storage tools kindly provided by the INAF Astronomical Archive (IA2) initiative (http://www.ia2.inaf.it). 

\section*{Data Availability}

Both the tracer data used for this work  \footnote{\hyperlink{https://owncloud.ia2.inaf.it/index.php/s/zcskRy1bryHN62i}{https://owncloud.ia2.inaf.it/index.php/s/zcskRy1bryHN62i}} and the IDL code used to evolve their energy spectra  \footnote{\href{https://github.com/FrancoVazza/IDL_FP}{https://github.com/FrancoVazza/IDL$\_$FP}}  are publicly available. The \textsc{ENZO} code used to produce the cosmological simulation is also publicly available \footnote{\url{enzo-project.org}}.



\bibliographystyle{mnras}
\bibliography{franco3}



\appendix

\section{Averaged MS scenario radio emission}
\label{sec:appendix}
In this section, we describe in more details the radio emission obtained from the family-averaged analysis introduced in Sec. \ref{sec:families}.

\subsection{Relic A}
\label{sec:relicA}
\begin{figure}
    \centering
    \includegraphics[width=\columnwidth]{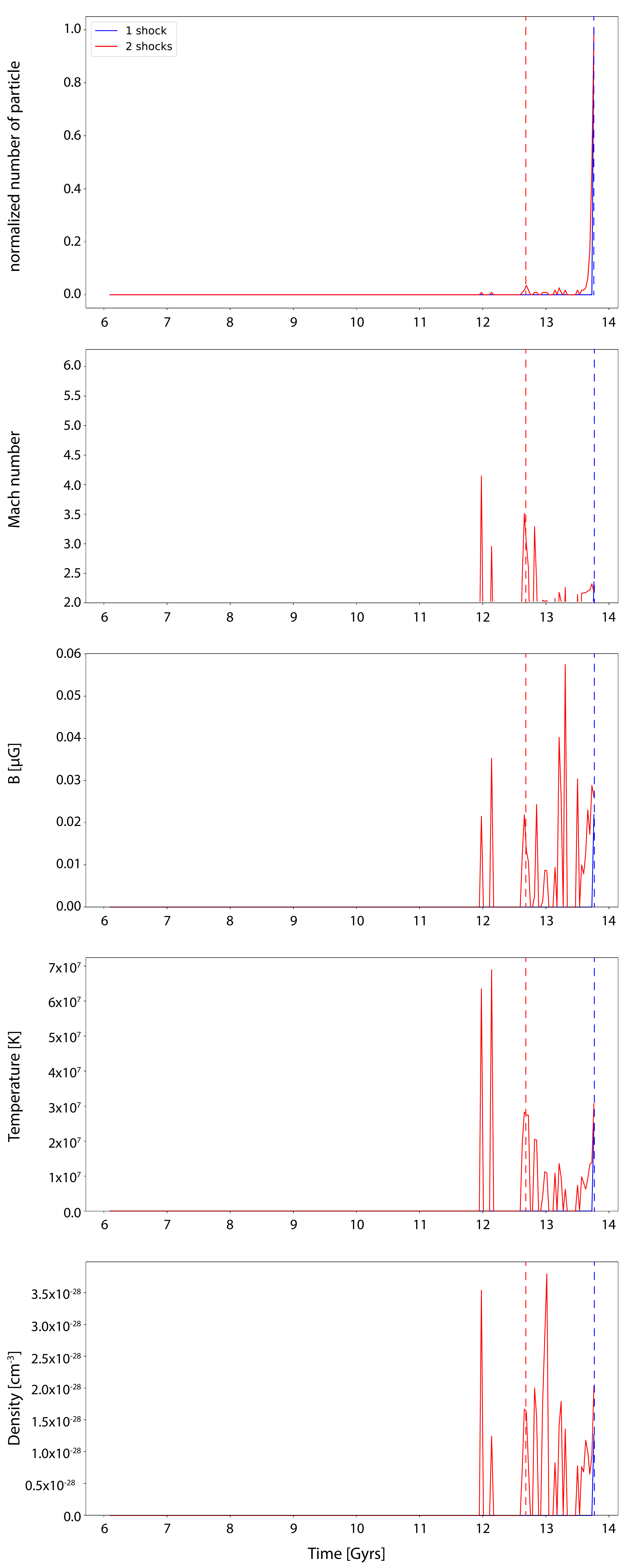}
    \caption{From top to bottom: time evolution of the number of shocked tracers, the averaged Mach number, the averaged magnetic field, the averaged temperature and the averaged density for both Family 1 (blue) and Family 2 (red) for Relic A. The vertical dashed lines mark the shock times chosen to represent each family.}
    \label{fig:ave_A}
\end{figure}

Figure \ref{fig:ave_A} shows the time evolution of the number of shocked tracers, the averaged Mach number, the averaged magnetic field, the averaged temperature and the averaged density for both Family 1 (blue) and Family 2 (red) for Relic A.
The vertical dashed lines mark the shock times chosen to represent each family.

\begin{figure}
    \centering
    \includegraphics[width=\columnwidth]{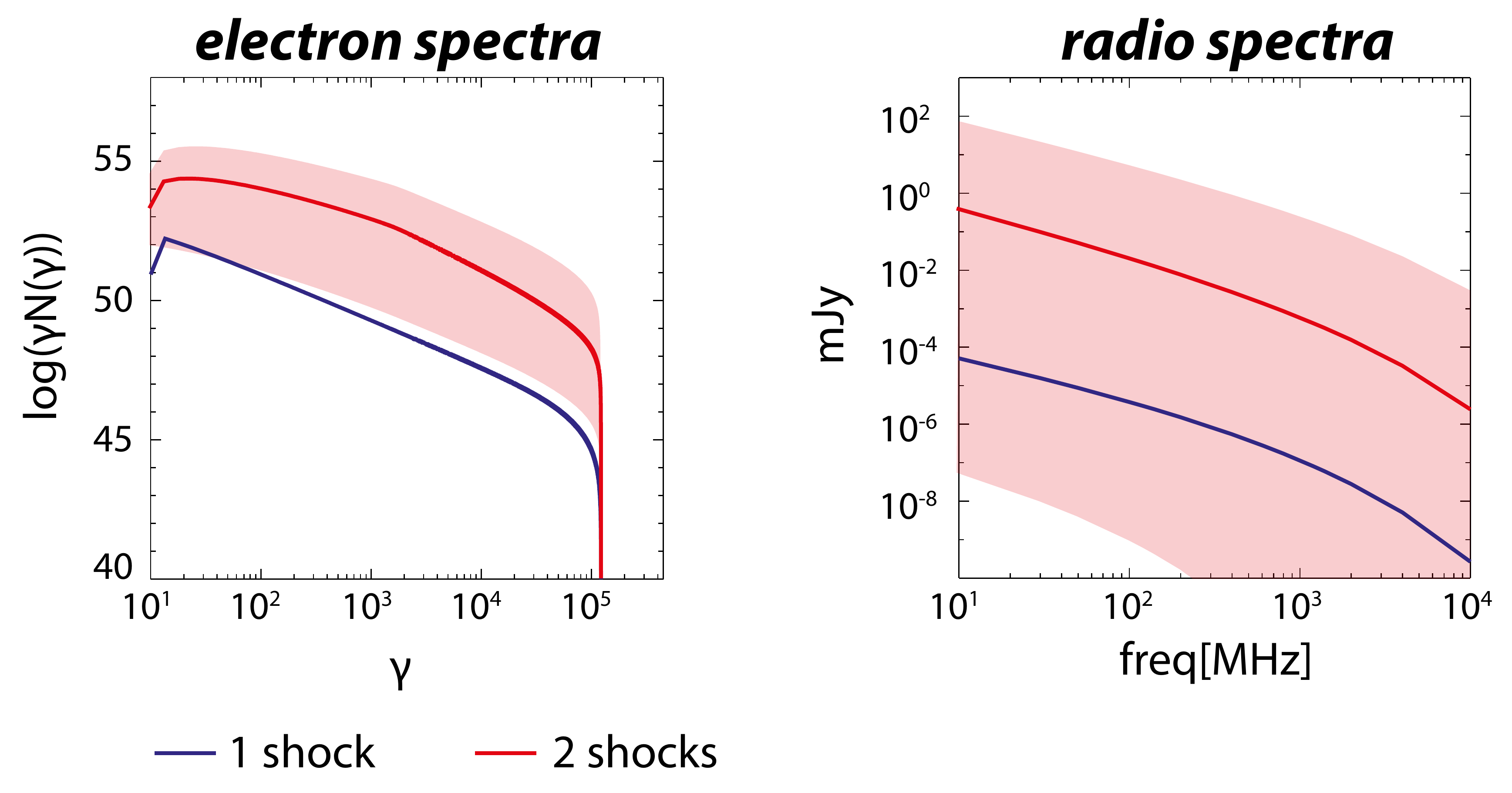}
    \caption{Electron energy spectra (left) and radio  spectra (right) at $t_{\mathrm{end}}$ for the population-averaged quantities for Family 1 (blue) and Family 2 (red) in Relic A. The wide red area shows the range of variability of Family 2 spectra using the $\pm\sigma$ values of the averaged quantities as input parameters.}
    \label{fig:spectra_A}
\end{figure}

Using the electron energy spectrum obtained at $t_{\mathrm{end}}$, we compute the radio spectrum associated for each family.
We numerically integrated the synchrotron emission from  tracers assuming the Jaffe Perola (JP) ageing model \citep[][]{1973A&A....26..423J}, and we placed our cluster at the approximate distance of the Coma cluster (i.e. $\approx 100 \rm ~Mpc$). 
In particular, we weight the spectra according to the relative population of the corresponding family.
Figure \ref{fig:spectra_A} shows the electron energy spectra (left) and radio  spectra (right) at $t_{\mathrm{end}}$ for the family-averaged quantities for Family 1 (blue) and Family 2 (red) in Relic A.
The wide red area in the plot shows the range of variability of the Family 2 spectra using the $\pm\sigma$ values of the averaged quantities as input parameters, where $\sigma$ is the standard deviation.
The standard deviation parameters are reported in Tab. \ref{tab:relicA_sigma}
\begin{table}
    \centering
    \begin{tabular}{c|c|c|c|c|c}
         & Time [Gyrs] & Mach & B [$\mu$G] & $\rho$ [g/cm$^{-3}$] & T [K]\\
    \hline
    Family 2 & & & & \\
    Shock 1 & 12.69 & $\pm$1.2 & $\pm$1.2e-02 & $\pm$5.2e-29 & $\pm$2.7e+07 \\
    Shock 2 & 13.76 & $\pm$0.4 & $\pm$2.1e-01 & $\pm$1.2e-28 & $\pm$1.0e+07
    \end{tabular}
    \caption{Standard deviation $\sigma$ associated to Family 2-averaged quantities in Relic A.}
    \label{tab:relicA_sigma}
\end{table}

To consider the radio emission variance associated with the family-averaged analysis, we present as case of study the variability for Family 2.
As we are going to present in this section, the results obtained from this analysis induced to proceed with the more detailed tracer integration analysis introduced in Sec. \ref{sec:integrated_radio}.
For this reason, we did not proceed to compute the averaged quantities for the other families.

We observe that the radio spectra obtained from the family-averaged analysis at $t_{\mathrm{end}}$ has similar slope for the two families.
However, the variation of the radio spectrum obtained using the standard deviation makes difficult to clarify if Family 2 has an higher emission than Family 1.

\subsection{Relic B}
\label{sec:relicB}
\begin{figure}
    \centering
    \includegraphics[width=\columnwidth]{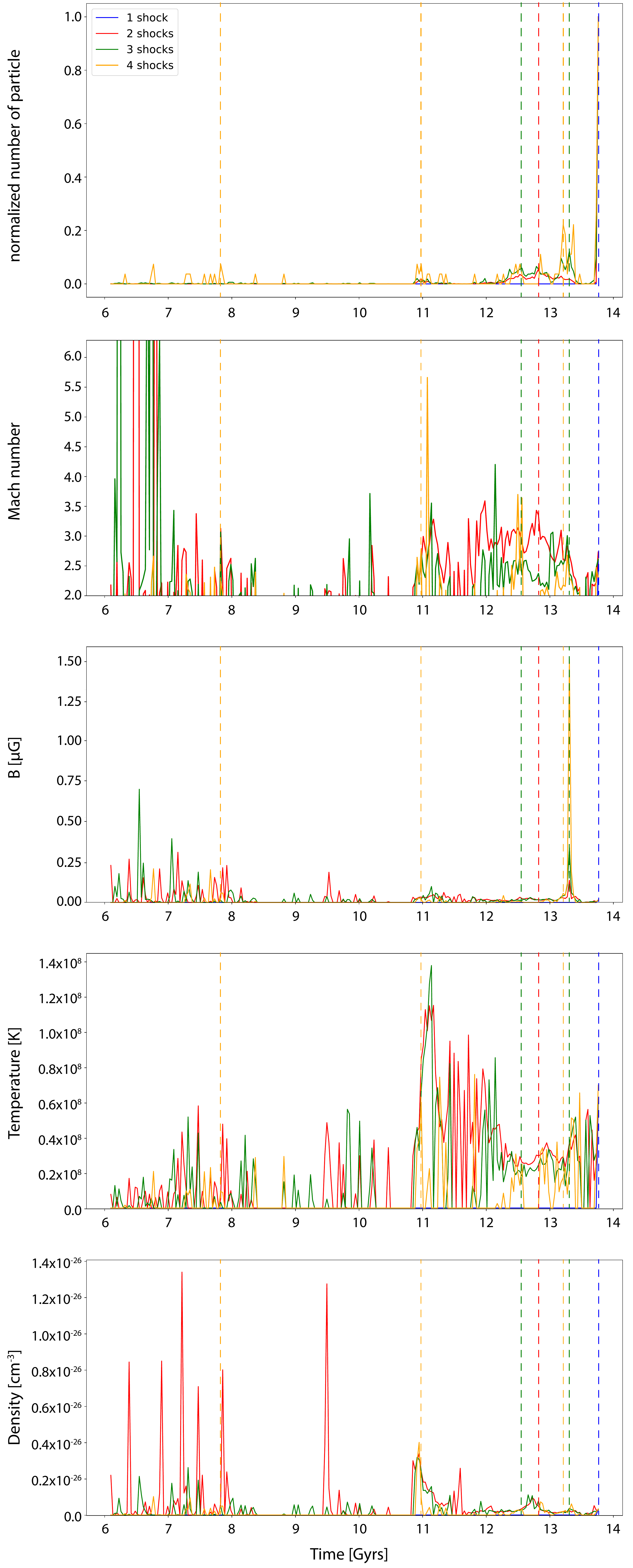}
    \caption{From top to bottom: time evolution of the number of shocked tracers, the averaged Mach number, the averaged magnetic field, the averaged temperature and the averaged density for Family 1 (blue), Family 2 (red), Family 3 (green), and Family 4 (orange) for Relic B. The vertical dashed lines mark the shock times chosen to represent each family.}
    \label{fig:ave_B}
\end{figure}

Figure \ref{fig:ave_B} shows the time evolution of the number of shocked tracers, the averaged Mach number, the averaged magnetic field, the averaged temperature and the averaged density for Family 1 (blue), Family 2 (red), Family 3 (green), and Family 4 (orange) for Relic B.
The vertical dashed lines mark the shock epochs chosen to represent each family in the population-averaged analysis.

\begin{figure}
    \centering
    \includegraphics[width=\columnwidth]{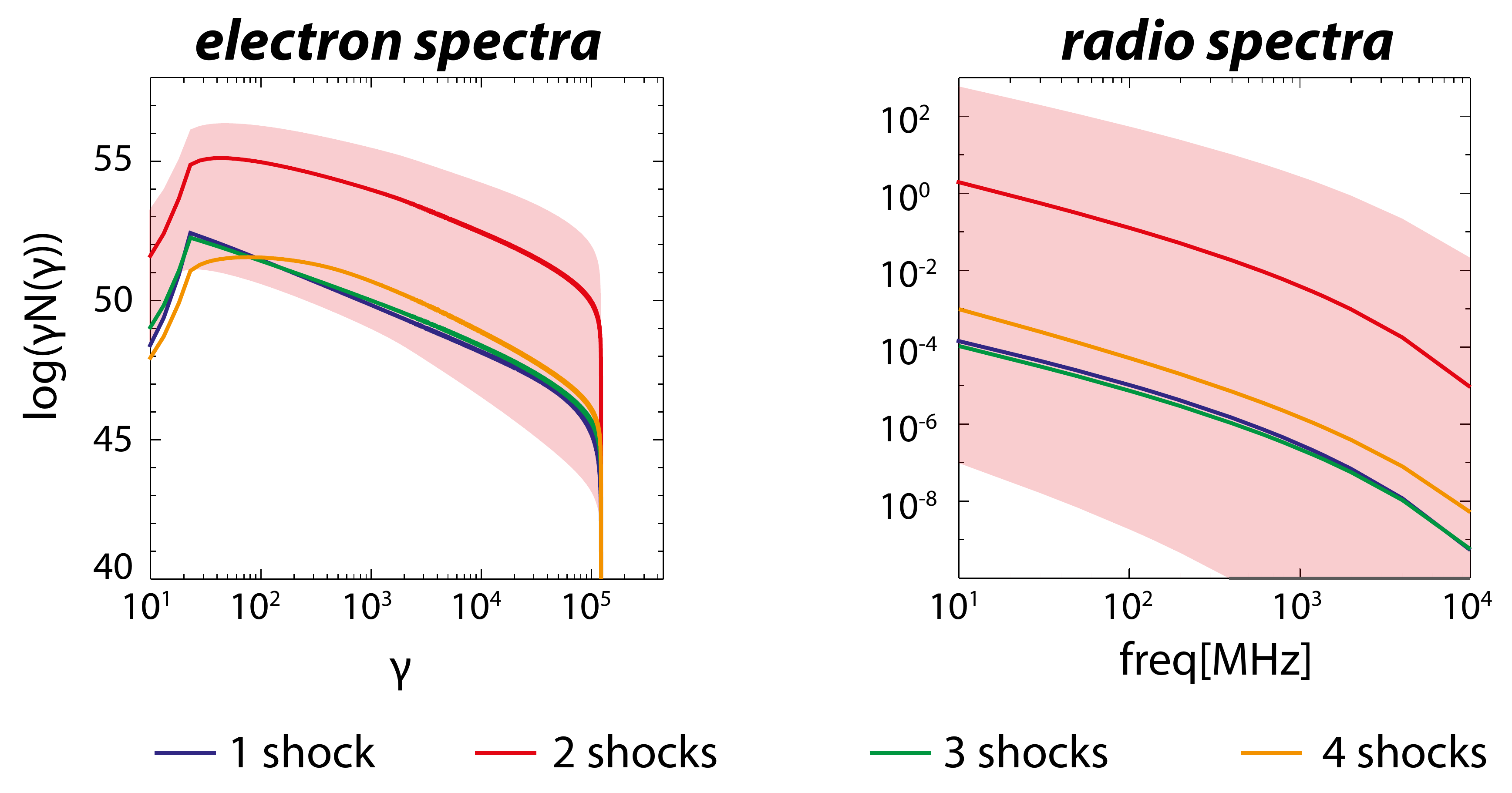}
    \caption{Electron energy spectra (left) and radio  spectra (right) at $t_{\mathrm{end}}$ for the averaged-quantities for Family 1 (blue), Family 2 (red), Family 3 (green), and Family 4 (orange) in Relic B. The wide red area shows the range of variability of Family 2 spectra using the $\pm\sigma$ values of the averaged quantities as input parameters.}
    \label{fig:spectra_B}
\end{figure}

Using the electron energy spectrum obtained at $t_{\mathrm{end}}$, we compute the radio spectrum associated with each family.
Figure \ref{fig:spectra_B} shows the electron energy spectra (left) and radio  spectra (right) at $t_{\mathrm{end}}$ for the averaged-quantities for Family 1 (blue), Family 2 (red), Family 3 (green), and Family 4 (orange) in Relic B.
The wide red area in the plot shows the range of variability of Family 2 spectra using, as input parameters, the $\pm\sigma$ values of the averaged quantities.
The standard deviation parameters are reported in Tab. \ref{tab:relicB_sigma}
\begin{table}
    \centering
    \begin{tabular}{c|c|c|c|c|c}
         & Time [Gyrs] & Mach & B [$\mu$G] & $\rho$ [g/cm$^{-3}$] & T [K]\\
    \hline
    Family 2 & & & & \\
    Shock 1 & 12.69 & $\pm$1.0 & $\pm$9.4e-02 & $\pm$5.0e-28 & $\pm$8.4e+06 \\
    Shock 2 & 13.76 & $\pm$0.9 & $\pm$1.1e-01 & $\pm$1.5e-28 & $\pm$2.3e+07
    \end{tabular}
    \caption{Standard deviation $\sigma$ associated to Family 2-averaged quantities in Relic B.}
    \label{tab:relicB_sigma}
\end{table}

We notice that the electron energy spectra for MS scenario families at the final stage of the simulation have a knee around  $\gamma\sim10^3$.
Since the radio emission is more relevant for high energies of the electron spectrum, the slope of the radio spectra is similar for all families, but Family 2 radio emission dominates over the others by a few orders of magnitude, reaching a peak of $\sim0.1$ mJy at $\sim100$ MHz.
However, as for Relic A, the variation of the radio spectrum obtained using the standard deviation makes difficult to clarify if Family 2 has an higher emission than the other families.

The results obtained from the family-averaged analysis of the two relics induced to proceed with the more detailed tracer integration analysis introduced in Sec. \ref{sec:integrated_radio}.
For this reason, we did not proceed to compute the averaged quantities for the other families.


\bsp	
\label{lastpage}
\end{document}